\documentclass[journal]{IEEEtran}
\usepackage{graphicx}
\usepackage{booktabs}
\usepackage{array}
\usepackage{tabularx}
\usepackage{placeins}
\usepackage{cite}
\usepackage{amsmath,amsfonts}
\usepackage{url}
\usepackage{caption}
\usepackage[hidelinks]{hyperref}
\usepackage[ruled,vlined,linesnumbered]{algorithm2e}
\SetKwInput{KwInit}{Initialize}
\usepackage{pifont}
\newcommand{\Description}[1]{}

\newif\ifshowmeshrefinement
\showmeshrefinementtrue

\makeatletter
\newcommand{\setupappendixfloats}{%
  \appendix
  \renewcommand{\thetable}{\thesection.\arabic{table}}%
  \renewcommand{\thefigure}{\thesection.\arabic{figure}}%
  \renewcommand{\thealgocf}{\thesection.\arabic{algocf}}%
  \setcounter{table}{0}%
  \setcounter{figure}{0}%
  \setcounter{algocf}{0}%
  \@addtoreset{table}{section}%
  \@addtoreset{figure}{section}%
  \@addtoreset{algocf}{section}%
}
\makeatother

\title{Topology-Aware Reinforcement Learning for CAD Decomposition toward All-Hexahedral Meshing}
\author{Jiaming Peng, Zhihao Zheng, Rui Wang, Qinglin Wang, Jing Xiao, Jie Liu, Na Lei, and Xinhai Chen%
\thanks{Jiaming Peng and Zhihao Zheng contributed equally to this work.}%
\thanks{Jiaming Peng, Qinglin Wang, Jing Xiao, Jie Liu, and Xinhai Chen are with the Laboratory of Digitizing Software for Frontier Equipment, the National Key Laboratory of Parallel and Distributed Computing, and the College of Computer Science and Technology, National University of Defense Technology, Changsha 410073, China.}%
\thanks{Zhihao Zheng is with the School of Data Science and Artificial Intelligence, Wenzhou University of Technology, Wenzhou 325000, China.}%
\thanks{Rui Wang is with the School of Mechanical Engineering and Intelligent Manufacturing, Ningbo University, Ningbo 315211, China.}%
\thanks{Na Lei is with the School of Software, Dalian University of Technology, Dalian, China.}%
\thanks{Corresponding author: Xinhai Chen.}}
\begin{document}
\maketitle

\begin{abstract}

Automatic hexahedral meshing for complex CAD models has long been regarded as the “holy grail” of mesh generation. The fundamental challenge arises from feature misalignment, where orthogonal or interacting features coexist, making direct hexahedral meshing infeasible and requiring decomposition into simpler, meshable blocks. While existing decomposition methods, including frame-field pipelines and learning-based techniques, have reduced manual intervention, they can remain sensitive to geometric and topological complexity and may fail to produce valid decompositions for challenging feature configurations. Specifically, their action spaces are typically restricted to surface-based splitting, which often yields invalid or unmeshable decompositions. To address these limitations, we propose DecompRL, a deep reinforcement learning framework for CAD decomposition toward hexahedral meshing. We design a set of topology-aware internal-splitting actions that leverage both global topological structure and local geometric property to effectively separate misaligned features. Furthermore, we introduce geometric rewards together with meshability feedback to guide the learning process toward valid decompositions and high-quality meshes. Extensive experiments on ABC, Fusion360, NIST, and DeepCAD show that DecompRL achieves the highest success rate among the evaluated learning-based methods on three of the four benchmarks, while maintaining competitive mesh quality. The measured LMFF-to-DecompRL runtime ratio reaches up to two orders of magnitude; this compares the full LMFF procedure with DecompRL decomposition alone, excluding subsequent mesh generation. The largest SR improvement is observed on Fusion360, where DecompRL reaches a success rate of 59.6\%.

\end{abstract}


\section{Introduction}

\begin{figure}[tbp]
\centering
\includegraphics[width=0.95\linewidth]{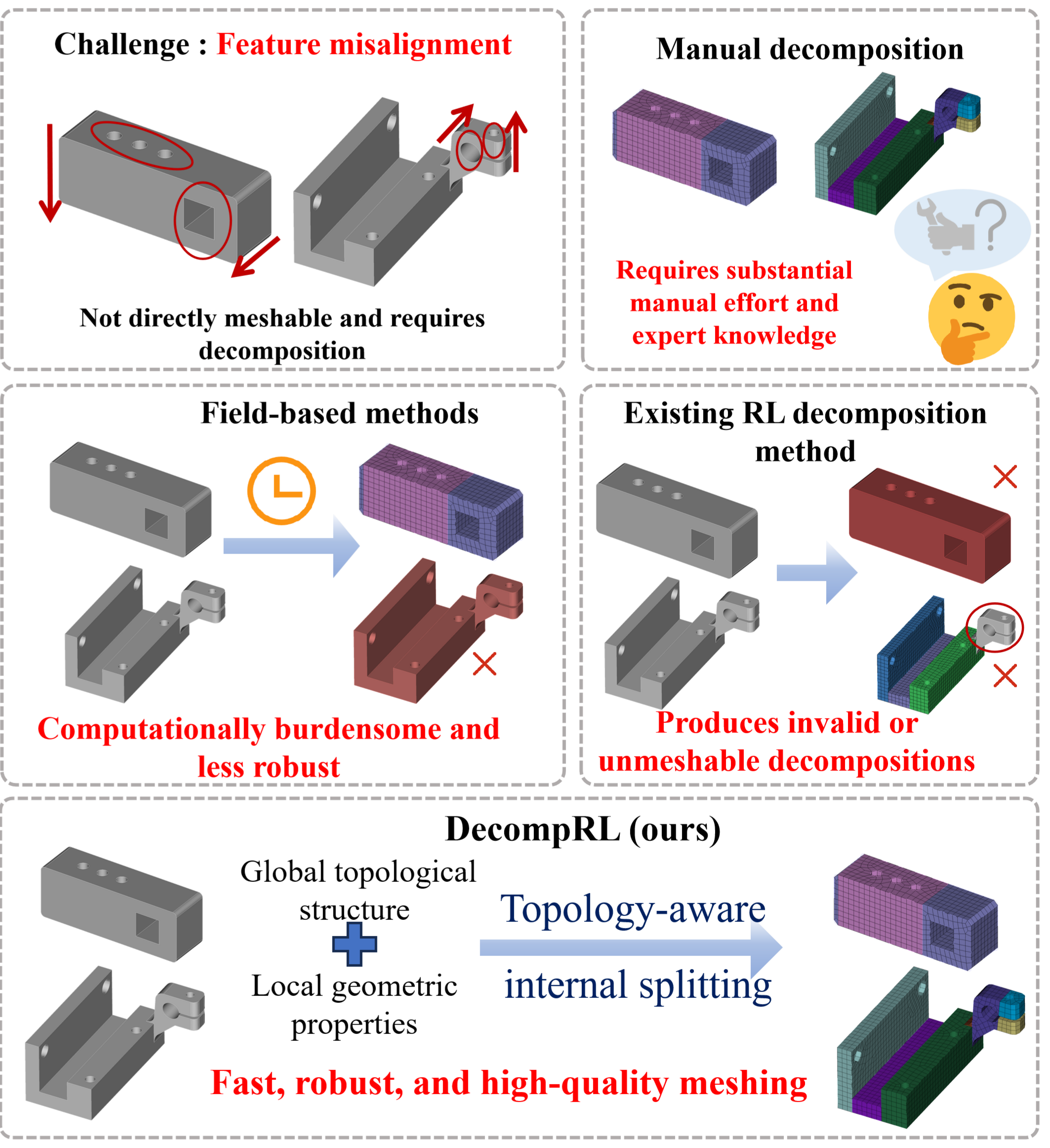}
\caption{Comparison of representative CAD decomposition and hexahedral meshing pipelines.}
\Description{Comparison of manual decomposition, field-based methods, existing RL decomposition, and DecompRL (ours).}
\label{fig:intro_comparison}
\end{figure}

\IEEEPARstart{H}{igh-quality} hexahedral meshing is highly desirable in many engineering workflows \cite{Pietroni2022HexMesh,Shepherd2008}. However, automatic hexahedral meshing for complex computer-aided design (CAD) models remains an open challenge in geometric modeling and simulation preprocessing \cite{Pietroni2022HexMesh,Shepherd2008}. A major bottleneck is feature misalignment: orthogonal or interacting features often coexist within the same CAD model and induce conflicting local blocking or sweep structures, so the model cannot be directly meshed as a whole. Fig.~\ref{fig:intro_comparison} shows a typical example, where features with orthogonal meshing directions coexist within a single CAD model, preventing the model from being directly meshed. Therefore, hexahedral meshing requires decomposing the CAD model into simpler sub-blocks that separate these misaligned features.

Representative approaches to decomposition-oriented hexahedral meshing can be organized by how the decomposition is constructed and selected: geometric/topological construction, field-guided construction, and learning-based decision-making. These directions may overlap and do not form an exhaustive taxonomy of hexahedral meshing algorithms. Manual, semi-automatic, and automatic instead describe the degree of human involvement in workflows that can employ these mechanisms. In practical engineering workflows, experienced users iteratively partition complex CAD models into meshable subvolumes. Prior work has noted that analysts may spend more than 50\% of the design-analysis cycle on decomposing complex models into simpler meshable blocks \cite{lu2017evaluation}. Classical construction methods exploit geometric and topological cues, including medial-axis or skeletal structures, feature recognition, and sweep- or block-based strategies \cite{Shepherd2008,Price1995MedialI,Price1997MedialII,Lai2000Sweeping,White2004CCSweep,Lu2001FeatureBased,sheffer2000virtual}.

Commercial CAD/CAE preprocessors combine interactive tools with automated decomposition and meshing capabilities. Ansys ICEM CFD supports structured blocking, block splitting and merging, geometric association, and automatic O-grid construction \cite{ANSYSICEM2025}. Altair HyperMesh provides mappability assessment, source/target detection, and tools for partitioning complex solids into simpler regions suitable for mapped meshing \cite{AltairHyperMesh2025}. Ansys MultiZone automatically decomposes geometry into mapped/sweepable and free regions, attempts to generate a pure hexahedral mesh, and permits alternative element types in free regions when the corresponding options are enabled \cite{ANSYSMultiZone2026}. These systems therefore already provide substantial automation for decomposition and hexahedral mesh preprocessing.

Within this workflow, DecompRL targets the construction and sequential selection of geometry-dependent decomposition operations. Geometric and topological analysis produces concrete candidates from the active B-Rep, and a learned policy selects an operation before continuing on the resulting child solids. Each partition changes the remaining geometry, available subsequent operations, and suitability for downstream meshing. DecompRL combines topology-aware internal-splitting candidates with learned action selection and sequencing to produce decompositions for subsequent structured-hexahedral meshing and verification. This role is complementary to existing CAD/CAE capabilities. Reducing manual trial-and-error is a potential engineering benefit; our experiments assess meshing outcomes, mesh quality, and computational runtime against the evaluated research baselines.

Field-based methods first compute a field on an input tetrahedral mesh and then extract dual surfaces, cutting sheets supported by stream surfaces, followed by hexahedral meshing \cite{Kowalski2014BlockStructured,Zheng2020DualSurfaces,Wang2020StreamSurfaceSheets}; Fig.~\ref{fig:intro_comparison} shows this pipeline schematically. However, these methods are often not robust. Constructing a valid frame field requires satisfying multiple geometric and topological constraints, which is challenging for complex models with misaligned features \cite{Huang2018SCOF,Brueckler2022Volume,Mitchell2023IIA}. In addition, field-based pipelines involve several coupled stages, so errors in early steps can easily propagate to later steps, making the overall process unstable and time-consuming.

Learning-based methods provide another direction, in which decomposition decisions are inferred from data \cite{Chen2022MGNet}. Among them, deep reinforcement learning (DRL) approaches formulate CAD model decomposition as a sequential decision-making problem, where each split produces multiple child sub-blocks and critically affects the topology and meshability of the remaining structure \cite{Pan2023Reinforcement,Zhang2025Block}. Within learning-based decomposition, we focus on reinforcement learning and investigate how a richer topology-aware action space can improve sequential CAD decomposition. However, the action space in current DRL methods is typically restricted to simple surface-based or planar splitting operations. Fig.~\ref{fig:intro_comparison} shows that these operations have limited capacity to resolve complex feature misalignments, often leaving the final sub-blocks unmeshable.

\begin{figure*}[t]
\centering
\includegraphics[width=0.9\textwidth]{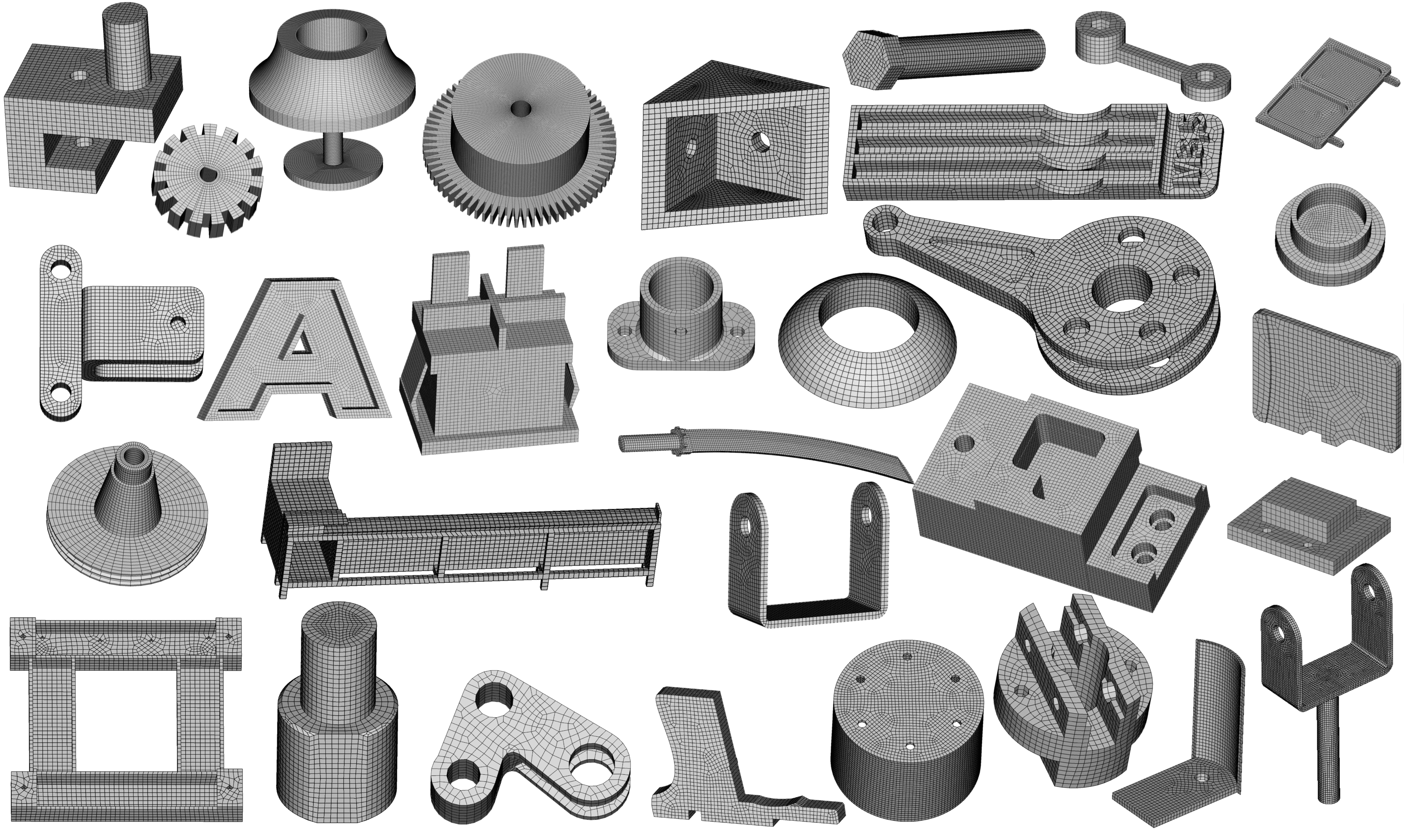}
\caption{Representative hexahedral meshing results produced by DecompRL across diverse CAD models.}
\Description{A wide overview image showing representative CAD model decompositions and the resulting hexahedral meshing results generated by DecompRL across diverse CAD models.}
\label{fig:overview_gallery}
\end{figure*}

To bridge these gaps, we propose DecompRL, a DRL framework for CAD model decomposition toward hexahedral meshing. DecompRL combines global topological structure and local geometric property to select topology-aware internal splitting actions that transform initially unmeshable CAD models into simpler, meshable sub-blocks. It further introduces a topology-aware action space that goes beyond surface-only splitting and incorporates meshability directly into the reward, so that policy learning is guided by whether a decomposition trajectory actually enables successful hexahedral meshing. Experimental results show that DecompRL achieves higher decomposition success rates on the evaluated CAD benchmarks, while maintaining competitive mesh quality and favorable computational efficiency. Fig.~\ref{fig:overview_gallery} shows representative outputs of the proposed method. Our contributions are:
\begin{itemize}
    \item \textbf{Problem formulation.} We formulate CAD model decomposition for hexahedral meshing as a feature misalignment problem, in which interacting geometric features are separated into simpler sub-blocks that support hexahedral meshing.
    \item \textbf{Topology-aware action space.} We design a topology-aware action space with candidate families spanning planar, symmetry, O-grid, feature gap, and rotation splits, providing diverse candidate operations for resolving different types of feature misalignment.
    \item \textbf{Reward with a meshability term.} We make hexahedral meshability an explicit part of the reward, encouraging the policy to prefer decomposition trajectories that more directly enable hexahedral meshing.
    \item \textbf{Evaluation.} We conduct extensive experiments on diverse public CAD benchmarks and show that DecompRL achieves higher decomposition success rates on three of four evaluated benchmarks while maintaining competitive hexahedral mesh quality.
\end{itemize}

\section{Related Work}
\label{sec:related_work}

\subsection{Classical Hexahedral Meshing and Decomposition Methods}

Classical hexahedral meshing methods can be roughly divided into direct generation techniques and those based on explicit decomposition and blocking. Cartesian-grid and octree-style methods are widely used in industrial preprocessing because they scale well and are comparatively easy to automate \cite{Aftosmis1998Cartesian,Pietroni2022HexMesh}. Another approach is dual-based extraction, which converts tetrahedral intermediates into hexahedral meshes \cite{Shepherd2008}. However, these methods exhibit weak alignment with detailed CAD features, boundary layers, and junction regions, leading to poor mesh quality that often requires substantial subsequent repair or optimization.

Another classical approach to hexahedral meshing first decomposes the CAD model into sweepable or block-structured subvolumes, and then generates the mesh. Medial axis and skeletal approaches derive block seeds from interior skeletons \cite{Price1995MedialI,Price1997MedialII}, while explicit blocking and sweep-based methods \cite{Lai2000Sweeping,White2004CCSweep,Lu2001FeatureBased,sheffer2000virtual} follow a more direct block construction strategy. These approaches can produce hexahedral meshes when the part admits a suitable sweep structure, but they rely heavily on feature recognition assumptions and are not robust.

\subsection{Frame-Field Methods for Hexahedral Meshing}

Field-based methods improve automation by constructing global orientation fields and then using them to guide segmentation, cutting, parameterization, or structured extraction. For example, Locally Meshable Frame Fields (LMFF) computes a frame field designed to better support hexahedral meshing \cite{liu2023locally}. Other field-based approaches extract intermediate structures, such as dual surfaces and cutting sheets, which are then used to guide decomposition and hexahedral meshing \cite{Kowalski2014BlockStructured,Zheng2020DualSurfaces,Wang2020StreamSurfaceSheets}.

Although these methods replace part of the manual blocking process, they are often not robust when handling complex CAD models with interacting features. Constructing a valid frame field requires satisfying multiple constraints, which is difficult in practice \cite{Huang2018SCOF}. Moreover, successful hexahedral meshing across the full pipeline depends on several coupled stages, including field construction, topology correction or surface extraction, and later discretization steps such as quantization or interval assignment \cite{Brueckler2022Volume,Mitchell2023IIA}. Errors in any of these stages can result in failure, and the multi-stage workflow also contributes to higher computational cost. In addition, directly generating a full hexahedral mesh makes subsequent local mesh refinement more challenging \cite{Pietroni2022HexMesh,2011Localized}.

\subsection{3D CAD Representation Learning}

For CAD model decomposition, representation learning concerns how the input CAD model is encoded. Early learning approaches often convert CAD models into voxel grids, which are convenient for 3D convolution but can blur thin features and weaken the face-edge adjacency relationships that matter for decomposition. More recent work operates directly on boundary representations (B-Reps), including BRepNet, BRepGAT, and UV-based surface representations \cite{Lambourne2021BRepNet,lim2023brepgat,jayaraman2021uv}. By representing CAD models as graphs, B-Reps preserve face and edge attributes together with their topological connectivity, making them a more suitable representation for CAD model encoding.

\subsection{Reinforcement Learning for Decomposition}

Reinforcement learning is well suited to CAD decomposition because the problem is sequential and long-horizon: each split changes the remaining geometry, the feasible future actions, and the likelihood of successful hexahedral meshing. RL has been used for decomposition problems in both 2D and 3D \cite{Pan2023Reinforcement,DiPrete2024Reinforcement,wu2025rlcad}. Most relevant to our setting, Zhang et al.~\cite{Zhang2025Block,Wu2024AAGNet} combine Soft Actor-Critic with imitation learning to predict split sequences for block decomposition toward hexahedral meshing. However, existing methods still rely on relatively limited split spaces, typically dominated by surface-based or planar operations, and on intermediate shaping rewards that are only weakly aligned with hexahedral meshability. As a result, they struggle to resolve feature misalignment in CAD models, often leaving decompositions that remain unmeshable.

\section{Method}
\label{sec:method}

\begin{figure*}[!tbp]
\centering
\includegraphics[width=\textwidth]{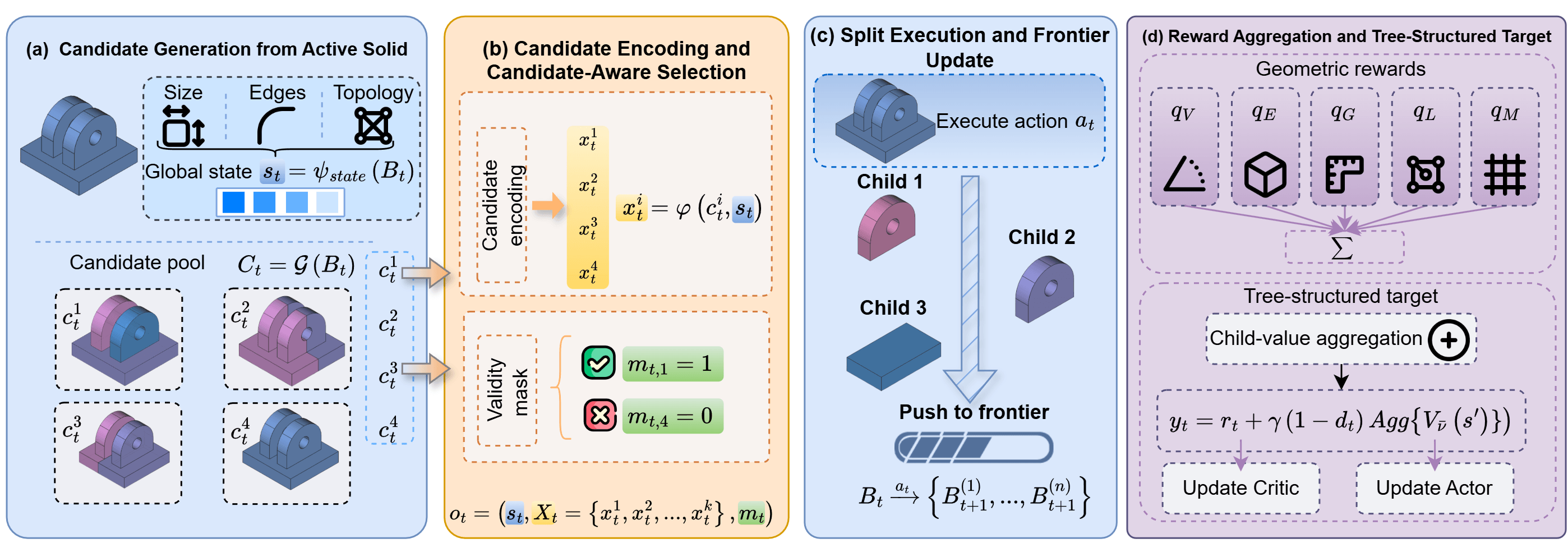}
\caption{Overview of DecompRL. (a) Starting from the active solid $B_t$, the environment constructs a typed candidate pool $C_t=\mathcal{G}(B_t)$ together with the state representation $s_t=\psi_{\text{state}}(B_t)$, which combines features extracted by a pretrained B-Rep encoder. (b) The generated candidates are encoded into the candidate set $X_t$, filtered by the validity mask $m_t$, and assembled into the policy observation $(s_t,X_t,m_t)$ for candidate selection. In the illustrated example, candidate $c_t^4$ does not yield a valid split, so its mask entry is set to $m_{t,4}=0$. (c) The selected geometric action $a_t$ is executed on $B_t$, producing child solids $\operatorname{Child}(B_t,a_t)$ that are pushed into the decomposition frontier for later decisions. (d) Reward terms and an aggregator over child values are combined to form the tree target used to update the actor and critics.}
\Description{A workflow diagram for the proposed method, showing B-Rep-based candidate construction, masked actor-critic decision making over candidate indices, geometric split execution with validity checks, and replay-based tree learning.}
\label{fig:method_overview}
\end{figure*}
DecompRL treats CAD model decomposition as a sequential decision process over an evolving decomposition frontier. At each step, the agent considers one active solid, generates candidate split operations from the B-Rep of that active solid, filters out infeasible candidates, and selects one valid split for execution. The selected split may produce multiple child solids, which are then inserted back into the frontier for subsequent decisions. This process continues until no further useful split remains or the step budget is reached. Fig.~\ref{fig:method_overview} shows the overall pipeline.

\begin{figure}[htbp]
\centering
\includegraphics[width=\linewidth]{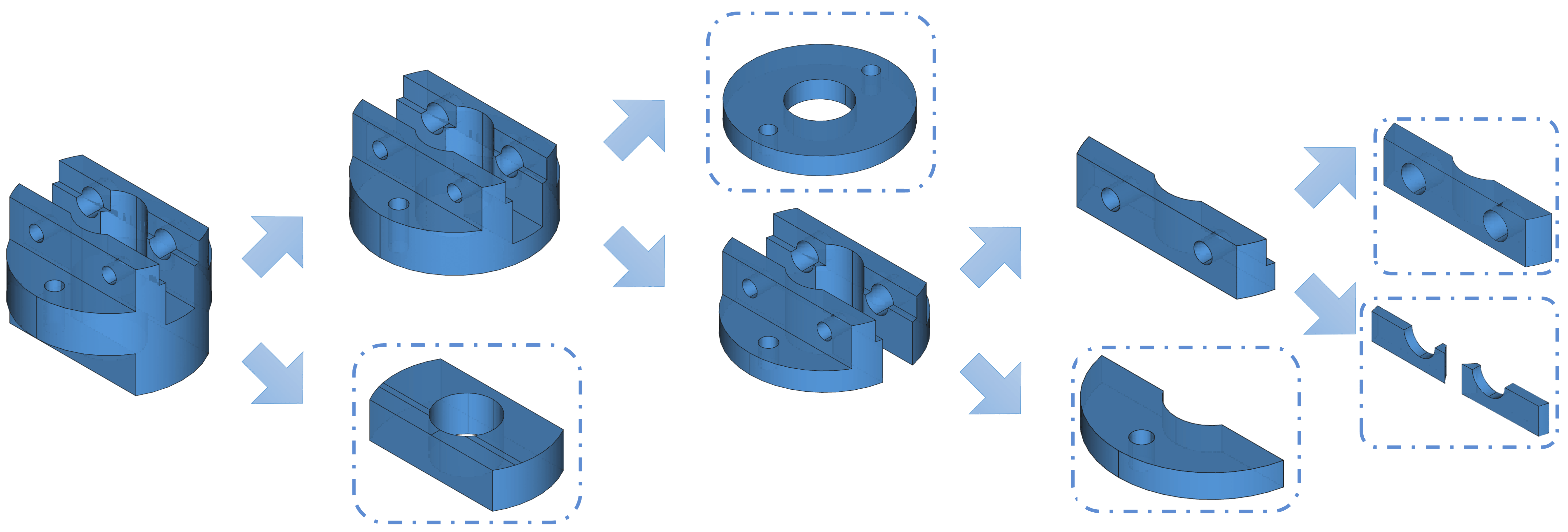}
\caption{Local frontier expansion in the decomposition tree. A selected split can generate multiple child solids, which reenter the decomposition frontier for later decisions.}
\Description{A local illustration of the decomposition frontier showing one CAD part being split into multiple child solids that branch into later decisions in a tree process.}
\label{fig:frontier_tree_expansion}
\end{figure}

\subsection{Problem Setup}
\label{sec:method_problem}

We formulate CAD decomposition as a decision-making process with branching transitions. Fig.~\ref{fig:frontier_tree_expansion} shows the local frontier expansion induced by a single split that produces multiple child solids. At decision step $t$, the environment considers one active solid, denoted by $B_t$, and generates a raw candidate pool $C_t=\{c_t^1,\ldots,c_t^{K_t}\}$ from that solid.

The policy observation combines three parts: a global state representation $s_t$ of the active solid, an encoded candidate set $X_t$ whose elements $x_t^i$ are computed from each candidate together with the global state context, and a binary validity mask $m_t \in \{0,1\}^{K_t}$ whose $i$-th entry indicates whether candidate $c_t^i$ remains admissible after filtering. The observation passed to the policy is therefore
\begin{equation}
o_t=(s_t,X_t,m_t).
\end{equation}
In the remainder of the paper, \emph{candidate pool} refers to the raw generated operations $C_t$, while \emph{candidate set} refers to the masked candidates presented to the policy after feasibility filtering and encoding.

Given this observation, the policy selects one valid candidate index $i_t \sim \pi_\theta(\cdot \mid o_t)$, which is then mapped to the executed geometric action $a_t $. A valid split under $a_t$ may produce multiple child solids and these children are inserted into the decomposition frontier and become future decision targets. An episode terminates when no splittable node remains or the step budget is reached.

\subsection{Action Space and Candidate Generation}
\label{sec:method_action}

Our action space consists of geometric split candidates constructed from the active solid rather than a fixed list of abstract actions. At decision step $t$, candidate generation produces the raw pool
\begin{equation}
C_t=\mathcal{G}(B_t)=\bigcup_k C_t^{(k)},
\end{equation}
where $\mathcal{G}(\cdot)$ denotes the candidate generation procedure and $C_t^{(k)}$ denotes the subset produced by candidate family $k$. The raw candidate pool $C_t$ is then formed by merging these family-specific candidates before feasibility filtering.

\begin{figure}[htbp]
\centering
\begin{minipage}[t]{0.48\linewidth}
\centering
\includegraphics[width=\linewidth]{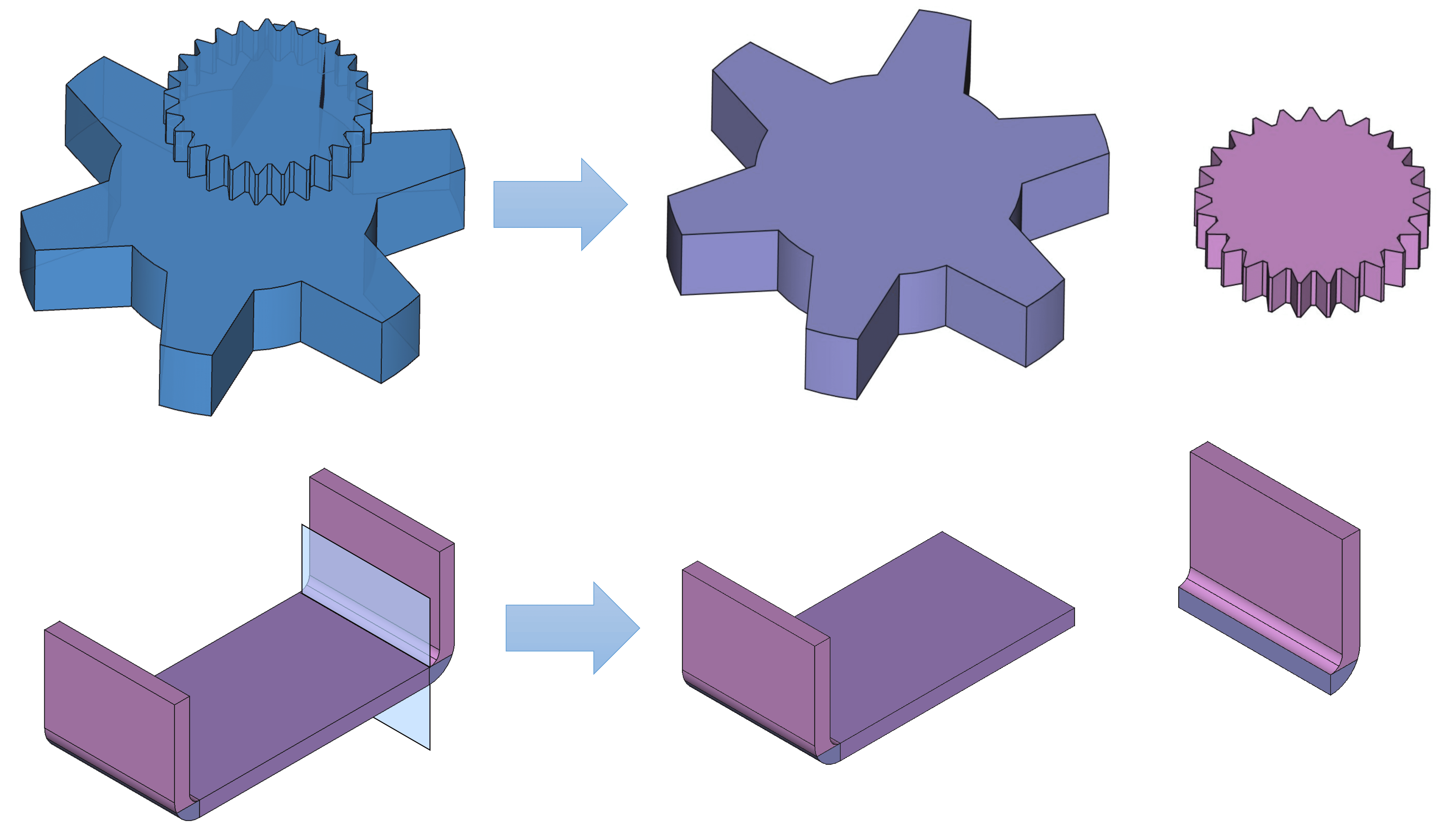}
\\[-2pt]\small (a) Planar split
\end{minipage}\hfill
\begin{minipage}[t]{0.48\linewidth}
\centering
\includegraphics[width=\linewidth]{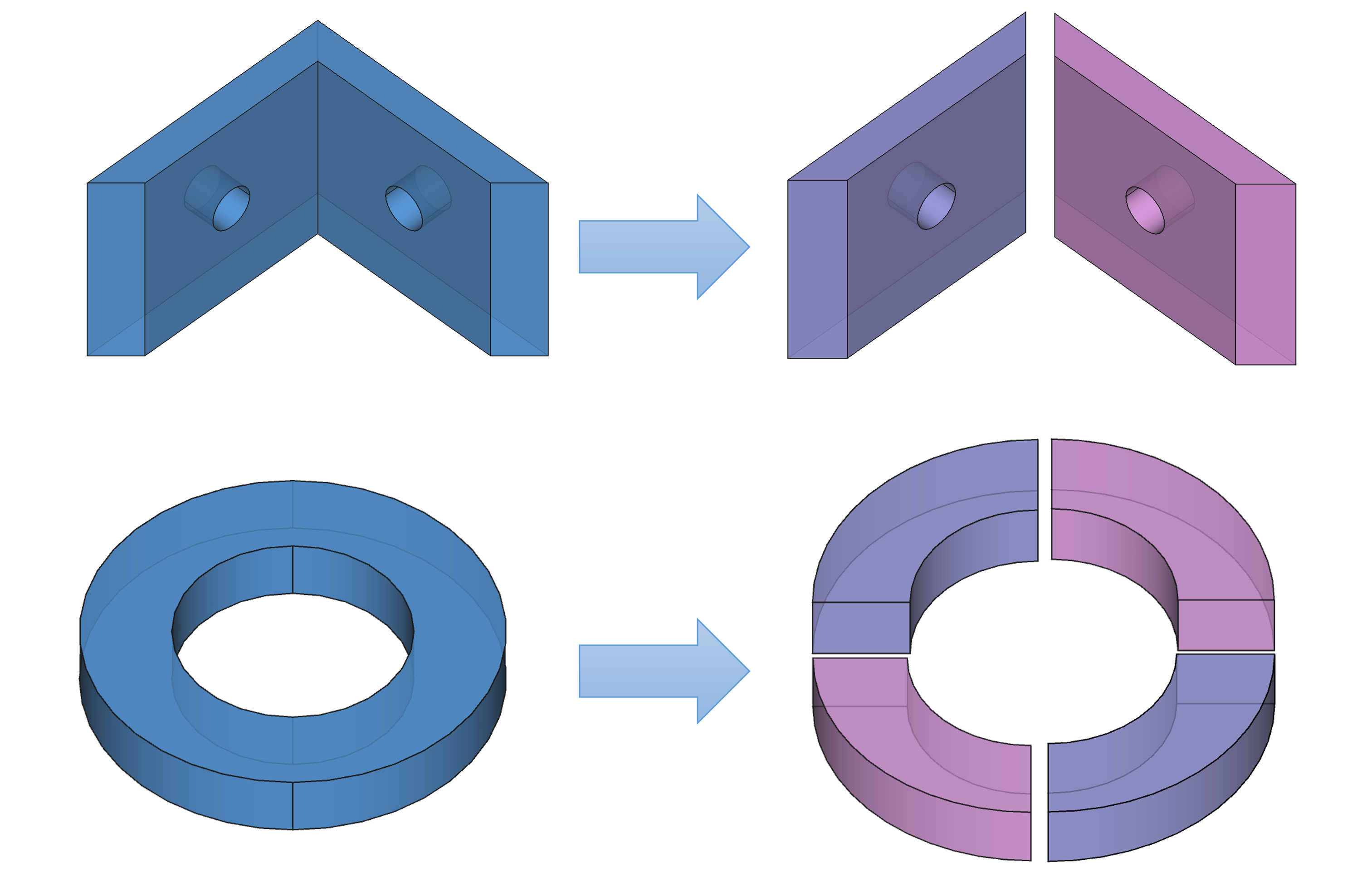}
\\[-2pt]\small (b) Symmetry split
\end{minipage}

\vspace{3pt}

\begin{minipage}[t]{0.48\linewidth}
\centering
\includegraphics[width=\linewidth]{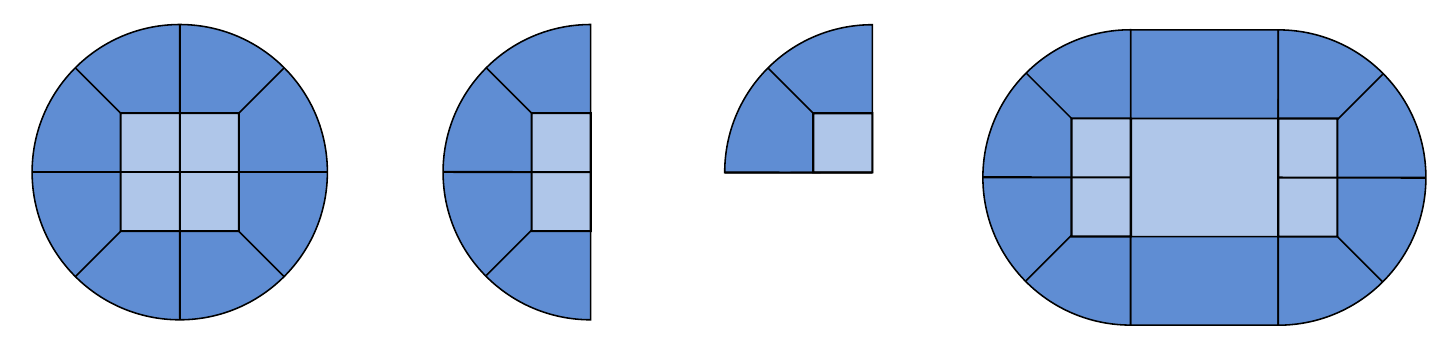}
\\[-2pt]\small (c) O-grid split
\end{minipage}\hfill
\begin{minipage}[t]{0.48\linewidth}
\centering
\includegraphics[width=\linewidth]{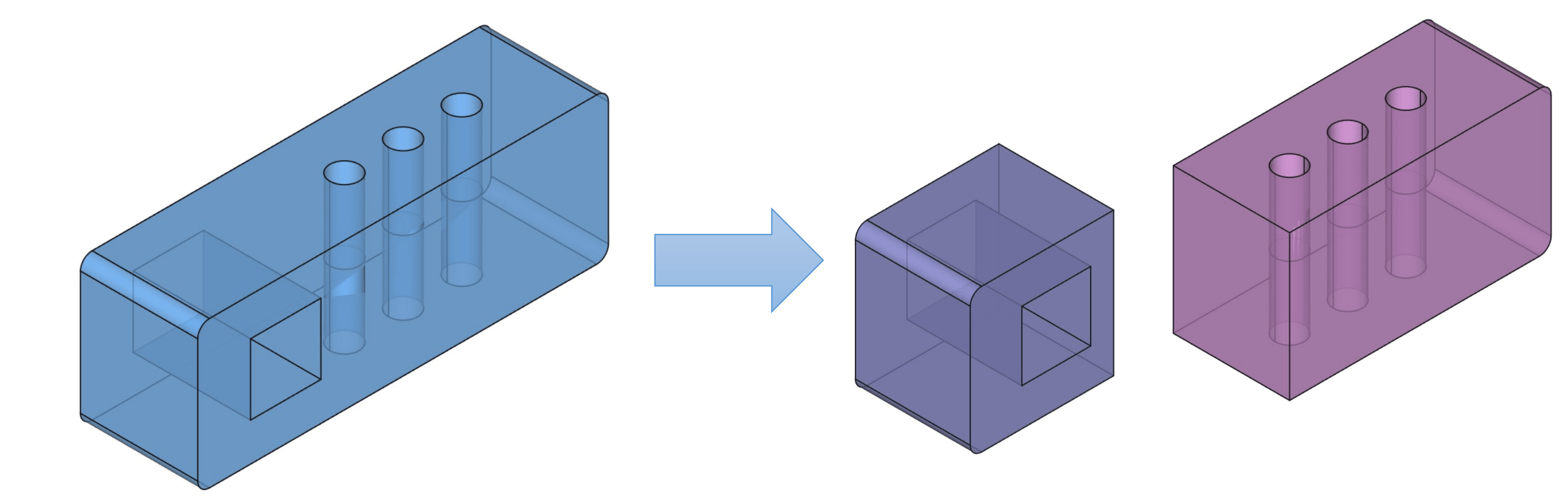}
\\[-2pt]\small (d) Feature gap split
\end{minipage}

\vspace{3pt}

\begin{minipage}[t]{0.58\linewidth}
\centering
\includegraphics[width=\linewidth]{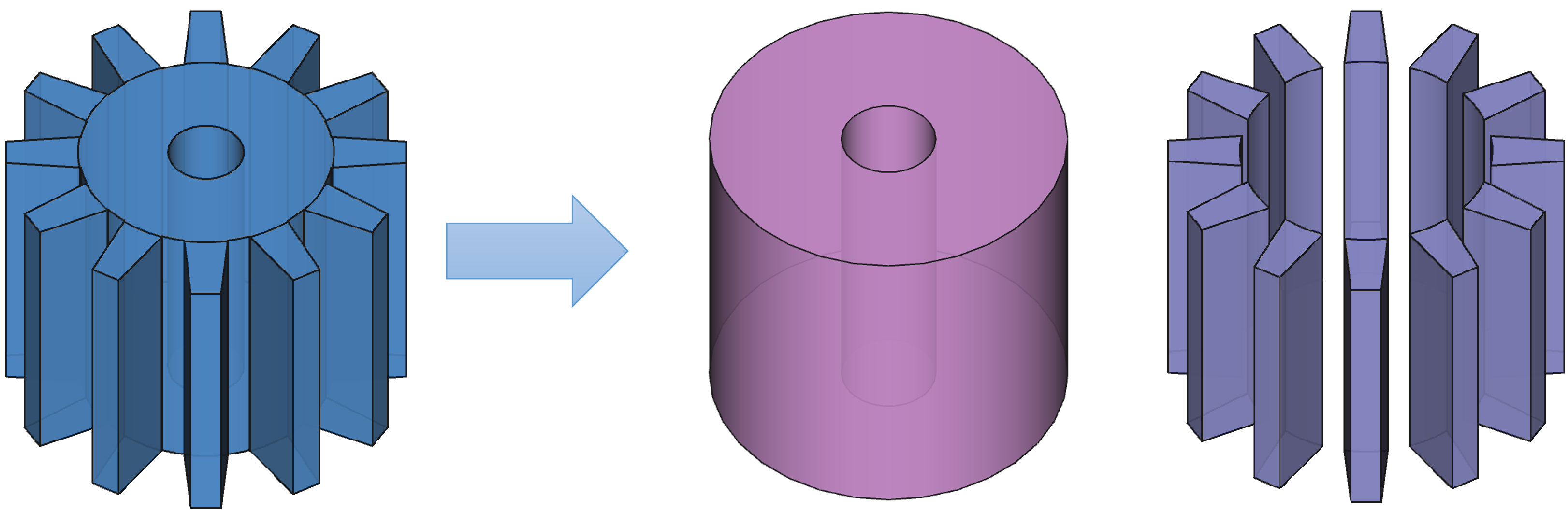}
\\[-2pt]\small (e) Rotation split
\end{minipage}
\caption{Planar, symmetry, O-grid, feature gap, and rotation splits.}
\Description{Five representative split types used by the policy: planar, symmetry, O-grid, feature gap, and rotation splits.}
\label{fig:action_primitives_overview}
\end{figure}

The candidate families are chosen to cover different blocking conflicts commonly encountered in hexahedral meshing, including planar, symmetry, O-grid, feature gap, and rotation splits. Here a candidate family denotes an operator class; at each decision step, geometry-specific proposal procedures instantiate these families on the current active solid $B_t$ to form a solid-specific raw candidate set $\mathcal{C}_t$. A planar split cuts the CAD model with a plane, which may coincide with a surface face or be an interior plane defined by three points, as shown in Fig.~\ref{fig:action_primitives_overview}(a). A symmetry split partitions the model along its symmetry plane. It is well suited for symmetric blocks, especially ring-like structures, as shown in Fig.~\ref{fig:action_primitives_overview}(b). O-grid splits improve mesh quality for circular structures and are particularly effective for resolving internal conflicts in intersecting cylindrical structures, such as T-junctions. They are also suitable for decomposing cylindrical and spherical blocks, as shown in Fig.~\ref{fig:action_primitives_overview}(c). A feature-gap split can be viewed as a special case of planar splitting. It separates nearby misaligned features by placing the cutting plane at the intermediate plane between them, as shown in Fig.~\ref{fig:action_primitives_overview}(d). The rotation split follows a cylindrical surface and separates the features on that surface from the cylindrical body, as shown in Fig.~\ref{fig:action_primitives_overview}(e). It is particularly effective for resolving sweep direction conflicts in ring-like structures. Together, these families provide complementary coverage for different forms of feature misalignment.

\begin{figure}[!t]
\centering
\includegraphics[width=\columnwidth]{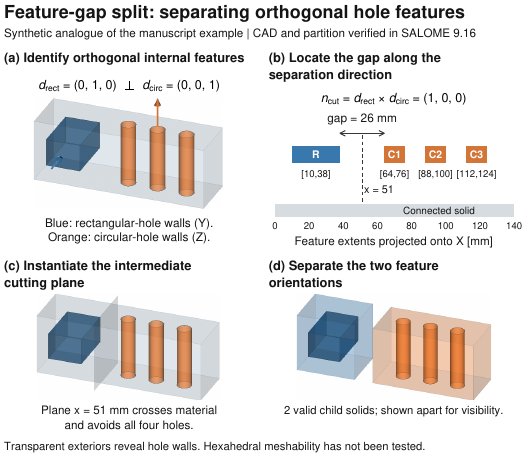}
\caption{Feature-gap candidate construction on a synthetic example. Orthogonal rectangular- and cylindrical-hole features are grouped, their extents are projected along the separation direction, and a plane is instantiated at the gap midpoint. Transparent exteriors reveal the hole walls. The example illustrates geometric partitioning; hexahedral meshability is not evaluated here.}
\Description{A four-panel illustration of feature identification, projected gap detection, midpoint cutting-plane construction, and the resulting two child solids.}
\label{fig:feature_gap_construction}
\end{figure}

\textbf{Feature-gap construction.} Local feature walls are grouped using opening relationships and topological adjacency, excluding common exterior faces that would otherwise connect unrelated features. Their projected extents are examined along geometrically determined separation directions. Overlapping intervals of all identified features are merged, so an intervening feature is not overlooked. A cutting plane is placed at the midpoint of a separation band satisfying the scale-dependent minimum-gap requirement, with its normal aligned with the separation direction. Candidates undergo geometric validity checks and deduplication, and at most five are retained for policy selection. Figure~\ref{fig:feature_gap_construction} illustrates this procedure for a transverse rectangular through-hole and vertical cylindrical through-holes: the midpoint plane crosses the connecting material while avoiding the hole features.

Before action selection, the raw candidate pool is filtered by feasibility checks to remove invalid or degenerate operations. Feasibility filtering is summarized by a binary mask
\begin{equation}
m_{t,i}=
\begin{cases}
1, & c_t^i \text{ passes the enabled checks},\\
0, & \text{otherwise},
\end{cases}
\end{equation}
so that only feasible candidates are passed to the policy. Here $m_{t,i}=1$ if candidate $c_t^i$ passes all geometric validity checks (non-degenerate volume, valid topology, and non-self-intersecting result), and $0$ otherwise. The checks remove candidates that violate geometric validity, minimum size or thickness constraints. As a result, the policy scores a compact set of feasible and structurally meaningful split options.

\subsection{State and Candidate Encoding}
\label{sec:method_state}

To make candidate selection depend on both global topological structure and local geometric property, we encode the observation with two complementary branches: one for the active solid and one for each candidate split. The state branch represents the active solid through a global representation derived from the boundary representation (B-Rep):
\begin{equation}
s_t = \psi_{\text{state}}(B_t),
\end{equation}
where $\psi_{\text{state}}(\cdot)$ combines a graph-level embedding from the AAGNet B-Rep graph encoder~\cite{Wu2024AAGNet} with an explicit 9-dimensional state descriptor of the active solid. The graph-level embedding provides learned B-Rep geometric/topological context from face attributes and face-adjacency relations, whereas the 9D state descriptor supplies fixed-size global geometric statistics. We initialize the AAGNet backbone from the released MFInstSeg-pretrained weights and keep it frozen during RL training; its machining-feature recognition heads are not used as a trainable component of DecompRL.

Each candidate split is encoded as
\begin{equation}
x_t^i = \varphi(c_t^i, s_t),
\end{equation}
where $\varphi(\cdot)$ combines a fixed-dimensional explicit candidate descriptor with candidate-associated local geometric/B-Rep context. The explicit descriptor records the candidate type, orientation, normalized relative position, and candidate-specific geometric attributes, while the local context describes the region of the active solid relevant to that candidate. The encoded candidate set is
\begin{equation}
X_t=\{x_t^1,\ldots,x_t^{K_t}\}.
\end{equation}
The state representation $s_t$ provides the global geometric/topological context shared by all candidates of the current active solid, whereas each $x_t^i$ describes one instantiated candidate through its explicit candidate descriptor together with candidate-associated local geometric/B-Rep context. The actor therefore evaluates every candidate conditionally on the same global state. Different candidate families retain a common explicit descriptor layout; entries corresponding to non-applicable candidate-specific attributes are set to zero.

The actor evaluates each candidate using the global state encoding and its candidate encoding, and the resulting scores are normalized with a feasibility mask:
\begin{equation}
\pi_\theta(i\mid s_t, X_t, m_t)
=
\mathrm{MaskedSoftmax}\big(\{f_\theta(s_t,x_t^j)\}_{j=1}^{K_t},\, m_t\big)_i,
\end{equation}
where $f_\theta$ is the actor scoring network. Infeasible candidates receive zero probability.

The policy action at step $t$ is the selection of one feasible instantiated candidate $c_t^i$ from the solid-specific candidate set, rather than the selection of an action-family label alone. Each candidate already specifies the geometry needed to instantiate its split, while the actor learns to rank these candidates under the current state. The critic evaluates the same candidate-conditioned decisions through twin Q-networks $Q_{\phi_1}(s_t,x_t^i)$ and $Q_{\phi_2}(s_t,x_t^i)$, taking their minimum to reduce value overestimation, and a separate state value network $V_\nu(s_t)$ with its target counterpart $V_{\bar{\nu}}(s_t)$ supports the tree backup described in Section~\ref{sec:method_training}. During training, the candidate index $i_t$ is sampled from the masked policy, whereas during inference the highest-probability feasible candidate is selected.

\paragraph{Feature dimensions}
In the reported configuration, the explicit \emph{9D state descriptor} included in $s_t$ contains the bounding-box scales $(b_x,b_y,b_z)$, diagonal length, volume, number of faces, number of edges, number of solids, and a binary multi-solid indicator $I_{\mathrm{multi}}=\mathbb{1}[N_{\mathrm{solid}}>1]$. This state-side descriptor summarizes global properties of the active solid and is distinct from the 9 candidate-specific geometric attributes defined below.
The explicit candidate descriptor in $x_t^i$ is 56-dimensional, consisting of a 41-dimensional one-hot encoding for candidate type, a 3D axis or surface-normal descriptor, a 3D normalized relative-position descriptor, and 9 candidate-specific attributes. These are the solid-cutter indicator, normalized candidate-associated length, generator-provided score, gap-family indicator, normalized gap size, normalized associated area, normalized radius, and two sweep-related slots. With $D$ denoting the active-solid bounding-box diagonal, length, gap size, area, and radius are encoded as $\ell_c/D$, $g/D$, $A/D^2$, and $r/D$, respectively. The solid-cutter indicator is 1 when a candidate is marked as using a volumetric cutting tool and 0 otherwise; it does not indicate execution success. The score previously termed confidence is supplied by the generator and is not a calibrated success probability. The gap-family indicator identifies gap-type candidates. Attributes not supplied by a generator are encoded as zero.

\subsection{Reward Design}
\label{sec:method_reward}

We design the reward to favor decompositions that remain geometrically regular and better aligned with hexahedral meshing. The geometric reward is composed of five terms (Eq.~\ref{eq:reward_composite}):
\begin{equation}
r_t^{\text{geo}}=\sum_{k\in\{V,E,G,L,M\}} \bar{w}_k\,q_{k,t}, \qquad
\bar{w}_k=\frac{w_k}{\sum_j w_j},
\label{eq:reward_composite}
\end{equation}
where the five reward terms are defined over the child blocks generated by the selected split.

The first two terms, $q_{V,t}$ and $q_{E,t}$, measure angular regularity. Specifically, $q_{V,t}$ evaluates face-corner angle quality: the in-face angle formed by adjacent boundary directions in the local face plane/tangent plane; it is not a 3D solid angle. The term $q_{E,t}$ evaluates the dihedral angle between adjacent face normals along a shared edge. For curved edges where the dihedral varies spatially, we evaluate the adjacent face normals at a representative point near the curve center and use their included angle as a local edge measure, rather than integrating or taking the worst angle over the entire edge. In each case, the reward combines the mean angle quality of the generated child blocks with a relative quality-preservation term that compares scalar child and parent angle-quality references. This comparison does not measure geometric or topological similarity between the child and parent blocks. For these two terms, we use the shared template (Eq.~\ref{eq:reward_angle_template})
\begin{equation}
\mathcal{Q}(\mu,c,p)=0.5\,\mu+0.5\,\min\!\left(\left(\frac{c}{p+\epsilon}\right)^2,1\right),
\label{eq:reward_angle_template}
\end{equation}
and instantiate (Eq.~\ref{eq:reward_qv_qe})
\begin{equation}
\begin{aligned}
q_{V,t}&=\mathcal{Q}(\alpha_{t}^{\mathrm{mean}},A_{t}^{\mathrm{child}},A_{t}^{\mathrm{par}}), \\
q_{E,t}&=\mathcal{Q}(\beta_{t}^{\mathrm{mean}},D_{t}^{\mathrm{child}},D_{t}^{\mathrm{par}}),
\end{aligned}
\label{eq:reward_qv_qe}
\end{equation}
where $\alpha_t^{\mathrm{mean}}$ and $\beta_t^{\mathrm{mean}}$ denote the mean face-corner angle and dihedral-angle quality scores over the generated child blocks, respectively. The terms $A_t^{\mathrm{child}}$ and $A_t^{\mathrm{par}}$ denote the child and parent face-corner angle-quality reference terms, while $D_t^{\mathrm{child}}$ and $D_t^{\mathrm{par}}$ denote the corresponding child and parent dihedral-angle quality reference terms. The preservation term is one-sided and saturates at 1 when the child reference is equal to or better than the parent reference; it only decreases when the split degrades the corresponding quality measure. Thus, structural changes such as those intentionally introduced by an O-grid operation are not penalized simply for differing from the parent.

The extreme-geometry term $q_{G,t}$ (not to be confused with the candidate generator $\mathcal{G}$) acts as a relative degeneracy guard. It discourages a split from introducing substantially worse short-edge, poor face-shape/aspect, or small-face-area conditions in the generated children (Eq.~\ref{eq:reward_qg}):
\begin{equation}
\begin{aligned}
q_{G,t} &= 0.25\,\eta_{t,\mathrm{len}}+0.5\,\eta_{t,\mathrm{aspect}}+0.25\,\eta_{t,\mathrm{area}}, \\
\eta_{t,x} &= \mathrm{clip}\!\left(\left(\frac{g_{t,x}^{\mathrm{child}}}{g_{t,x}^{\mathrm{par}}+\epsilon}\right)^2,0,1\right), \;
x\in\{\mathrm{len},\mathrm{aspect},\mathrm{area}\},
\end{aligned}
\label{eq:reward_qg}
\end{equation}
Here the compared quantities are scalar lower-bound geometric-quality measures rather than descriptors of the overall child shape or topology. The score reaches 1 when the corresponding child quality is preserved or improved relative to the parent and decreases only when the split creates a worse degeneracy.
The internal coefficients $0.25/0.5/0.25$ are fixed heuristic design weights rather than values obtained from test-set optimization or a dedicated hyperparameter search. The face-shape/aspect term receives the larger weight to emphasize narrow or strongly distorted faces, while the short-edge and small-area terms provide complementary indicators of scale-related degeneracy.

The term $q_{L,t}$ encourages simpler face-boundary topology by rewarding decompositions whose child blocks contain fewer B-Rep faces with internal face-boundary loops (Eq.~\ref{eq:reward_ql}). Here an internal face-boundary loop means an inner trimming loop of a face in addition to its outer boundary; it is not a direct count of 3D holes or tunnels in the solid:
\begin{equation}
q_{L,t}=\frac{1}{N_t}\sum_{i=1}^{N_t}\left(\frac{F_{t,\mathrm{no\_loop}}^{(i)}}{\max(F_{t,\mathrm{total}}^{(i)},1)}\right)^2,
\label{eq:reward_ql}
\end{equation}
where $N_t$ is the number of generated child blocks, $F_{t,\mathrm{no\_loop}}^{(i)}$ denotes the number of faces with no inner boundary loop in child block $i$, and $F_{t,\mathrm{total}}^{(i)}$ denotes the total number of faces in child block $i$.

Finally, the meshability term ($q_{M,t}$) provides a meshability-oriented signal through the mean of the child-block meshability indicators (Eq.~\ref{eq:reward_qm})
\begin{equation}
q_{M,t}=\frac{1}{N_t}\sum_{i=1}^{N_t} z_{t,i}^{\mathrm{mesh}},
\label{eq:reward_qm}
\end{equation}
where $z_{t,i}^{\mathrm{mesh}}\in\{0,1\}$ is the operational meshability indicator assigned to child block $B_{t,i}$, defined as
\begin{equation}
z_{t,i}^{\mathrm{mesh}}=
\mathbb{I}\!\left[
c_{\mathrm{sweep}}(B_{t,i})\vee
c_{\mathrm{map}}(B_{t,i})\vee
c_{\mathrm{submap}}(B_{t,i})
\right],
\label{eq:zmesh_def}
\end{equation}
where $c_{\mathrm{sweep}}$, $c_{\mathrm{map}}$, and $c_{\mathrm{submap}}$ denote the supported sweep, map, and submap compatibility checks, respectively. Thus, $q_{M,t}$ is the mean of binary meshability indicators at the child-block level and serves directly as a continuous feasibility signal for the set of child blocks. The positive weights $w_k$ are raw coefficients, while $\bar{w}_k$ denotes the normalized weight used in reward aggregation.
The child-level indicator is intentionally binary because Map, Submap, and Sweep are treated as alternative supported compatibility routes rather than ordered quality levels; satisfying several predicates does not automatically imply a higher-quality block than satisfying one valid predicate. The split-level score $q_{M,t}$ is nevertheless graded through the fraction of generated children that satisfy at least one supported compatibility condition.

\begin{figure}[!tbp]
\centering
\includegraphics[width=0.80\linewidth]{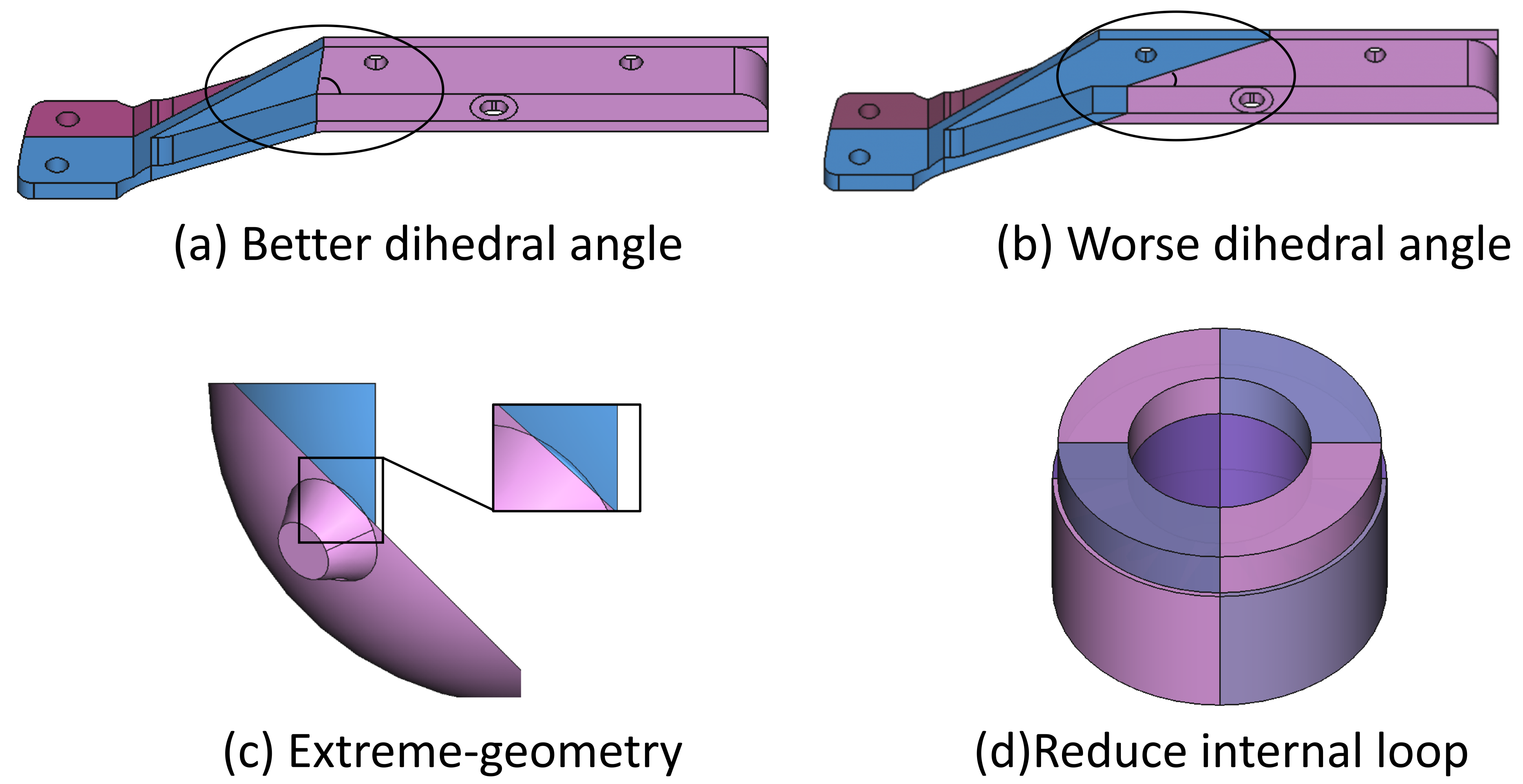}
\caption{Reward-guided geometric preferences. (a) and (b) compare dihedral-angle configurations for $q_E$. (c) shows an extreme geometry case penalized by $q_G$. (d) shows a case in which fewer internal loops improve $q_L$.}
\Description{A four part reward explanation figure for CAD model decomposition. The top row compares better and worse dihedral angle configurations, the lower left shows an extreme geometry case, and the lower right shows a shape with reduced internal loops.}
\label{fig:reward_terms_examples}
\end{figure}

Fig.~\ref{fig:reward_terms_examples} shows the geometric intuition behind the main reward terms: Fig.~\ref{fig:reward_terms_examples} (a)--(b) illustrate better and worse dihedral angle configurations for $q_E$, Fig.~\ref{fig:reward_terms_examples} (c) shows an extreme geometry case penalized by $q_G$, and Fig.~\ref{fig:reward_terms_examples} (d) shows how reducing internal loops improves $q_L$.

To handle invalid actions, stopping behavior, and depth bias in the decomposition tree, we further apply runtime shaping (Eq.~\ref{eq:runtime_reward_agg}):
\begin{equation}
r_t=\lambda_d^{h_t}\,r_t^{\text{base}}+\mathbb{I}_{\text{stop},t}\,b_{\text{stop}},
\label{eq:runtime_reward_agg}
\end{equation}
where
\begin{equation}
r_t^{\text{base}}=
\begin{cases}
-1.0, & \text{invalid split or empty children},\\
r_t^{\text{geo}}, & \text{otherwise}.
\end{cases}
\end{equation}
Here $h_t$ is the depth of the current node, $\mathbb{I}_{\text{stop},t}$ indicates whether stopping criteria are satisfied at step $t$, and $b_{\text{stop}}$ is the corresponding stop bonus. We use $\lambda_d=0.9$ as a tree-depth reward-shaping factor. The term $\lambda_d^{h_t}$ rescales the immediate geometric reward according to structural depth, so that obtaining the same local improvement only after additional recursive splits contributes less reward and unnecessarily deep decomposition trees are discouraged. This factor is distinct from the Bellman discount $\gamma=0.95$ used in Section~\ref{sec:method_training}: $\lambda_d$ acts on the current-node reward, whereas $\gamma$ discounts the aggregated future value of the generated child states. Thus, the two parameters serve different roles and are not interchangeable.
The stop flag and the terminal indicator are related but not identical: $\mathbb{I}_{\text{stop},t}=1$ marks a node that reaches a valid stopping state, whereas $d_t=1$ marks any terminal node, including those terminated by candidate exhaustion or split failure. Every stopped node is terminal ($\mathbb{I}_{\text{stop},t}=1\Rightarrow d_t=1$), but a node can be terminal without earning the stop bonus.

\subsubsection{Evaluation Criteria and Meshability Indicator}
\label{sec:eval_protocol}
This section clarifies the model-level evaluation criterion used in SR and the operational meshability indicator used in the reward meshability term $q_M$.

To guide the reinforcement learning agent toward valid geometric decompositions, we define an operational meshability indicator $z_{t,i}^{\mathrm{mesh}}\in\{0,1\}$ (Eq.~\ref{eq:zmesh_def}) for each decomposed child block $B_{t,i}$, evaluating its compatibility with the supported structured-hexahedral meshing families. Each compatibility function returns $1$ if the corresponding conditions hold and $0$ otherwise:
\begin{itemize}
    \item \textbf{Map compatibility $c_{\mathrm{map}}$}: the volume must admit a logical-hexahedral organization with six logical sides and eight logical corners. A logical side may consist of multiple geometric surfaces; opposite logical directions must admit compatible interval constraints so that a structured $i$-$j$-$k$ parametrization can be formed.
    \item \textbf{Submap compatibility $c_{\mathrm{submap}}$}: a volume that is not directly mappable as one logical hexahedron may still be accepted when its boundary admits a consistent local $i$-$j$-$k$ classification into mappable logical subregions. Boundary vertices are classified from the local interior-angle structure around nominal $90^\circ$ increments. The automatic classification uses a fixed fuzzy angular tolerance of $27^\circ$ for all datasets; this value is not tuned per dataset.
    \item \textbf{Sweep compatibility $c_{\mathrm{sweep}}$}: the volume must admit topologically compatible source and target surfaces connected by continuous linking surfaces. The linking surfaces must support Map/Submap-compatible structured discretization, and multiple logical source-to-target paths must have consistent total interval counts so that the source mesh can be propagated along one logical sweep direction.
\end{itemize}
These predicates are operational compatibility tests for the supported structured-hexahedral meshing routes rather than necessary-and-sufficient conditions for arbitrary hexahedral meshability. Consequently, local compatibility is used only as the meshability-oriented reward signal, while actual downstream meshing success is evaluated separately.

It is important to distinguish this intermediate reward signal from the final evaluation metric. While the block-level indicator $z_{t,i}^{\mathrm{mesh}}$ drives the optimization of $q_M$ during training, the model-level Success Rate (SR) reported in our evaluations is defined solely by successful completion of the final hexahedral meshing pipeline rather than by this operational indicator.
Successful generation is defined for the entire input CAD model, including all final leaf volumes. For each dataset, $\mathrm{SR}=100N_{\mathrm{success}}/200$ is reported as a percentage. A case ending without a complete hexahedral mesh, including incomplete volume meshing or errors or timeouts that prevent completion, is counted as a failure and remains in the denominator. Scaled-Jacobian statistics characterize the generated elements; the reported fraction with $J_s>0.8$ is not an additional threshold defining SR.
The current reward also does not directly optimize final element-wise hexahedral mesh quality. Such feedback could be incorporated in future work as a delayed terminal reward based on metrics such as scaled Jacobian, or through a learned mesh-quality surrogate. We keep it separate in the reported method because repeatedly generating a complete hexahedral mesh at intermediate decomposition steps would substantially increase training cost and couple the decomposition policy more tightly to the downstream mesher.

\subsubsection{Global Conformity Across Shared Block Interfaces}
\label{sec:global_conformity}

\begin{figure}[t]
\centering
\includegraphics[width=\columnwidth]{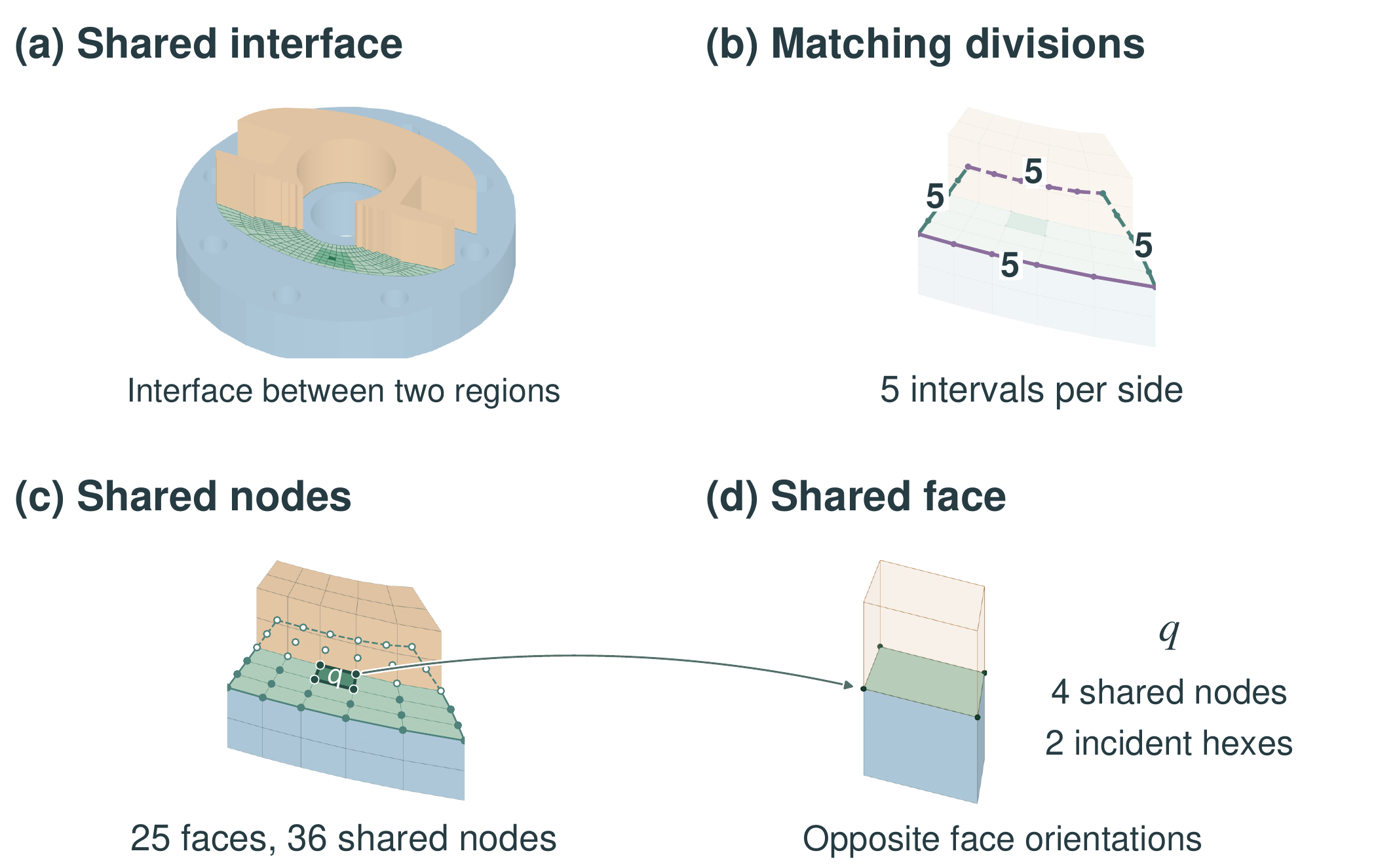}
\caption{Shared-interface conformity in the MAMBO M1 mesh generated with SALOME. (a) Selected interface patch. (b) Matching subdivisions with five intervals per side. (c) Shared patch mesh with 25 quadrilaterals and 36 nodes. (d) Two adjacent hexahedra share the four nodes of face $q$.}
\Description{Four panels show the M1 model, matching interface division constraints, shared nodes and face q on the same patch, and two neighboring hexahedra sharing that face.}
\label{fig:global_conformity}
\end{figure}

Block-level Map/Submap/Sweep compatibility is treated as a local operational meshability criterion, while global conformity is handled in a separate downstream meshing stage. After decomposition, the leaf volumes are assembled on shared CAD topology rather than meshed independently, so the final hexahedral mesh is generated on a common multi-block representation. Figure~\ref{fig:global_conformity} illustrates shared-interface conformity through a manually constructed MAMBO M1 verification example.

Coincident CAD interfaces are first subdivided into corresponding topological regions, and the matching surfaces, curves, and vertices are identified in the assembled model. For neighboring leaf volumes $V_i$ and $V_j$ sharing a two-dimensional interface $\Gamma_{ij}$, conformity requires
\begin{equation}
\mathcal M_i|_{\Gamma_{ij}}
=\mathcal M_j|_{\Gamma_{ij}}
=\mathcal M_{\Gamma_{ij}}.
\label{eq:shared_interface_mesh}
\end{equation}
Here $\mathcal M_i$ denotes the mesh of $V_i$, and $\mathcal M_{\Gamma_{ij}}$ is the common interface mesh. Equality means that both sides reference the same global node identifiers and quadrilateral connectivity. Each internal interface quadrilateral is incident on one hexahedron from each neighboring volume, with opposite boundary orientations. The interface mesh is generated once and reused by both volumes.

Interval compatibility also includes volume constraints. For a Sweep-compatible volume with fixed source, target, and logical edge classifications, let $n_e$ be the number of intervals on logical boundary curve $e\in E$. The signed interval sums along two source-to-target paths $P_r$ and $P_s$ must agree:
\begin{equation}
\sum_{e\in E}(w_{r,e}-w_{s,e})n_e=0.
\label{eq:sweep_path_intervals}
\end{equation}
The path coefficient $w_{r,e}\in\{-1,0,1\}$ is $+1$ or $-1$ for traversal along or against the logical sweep direction, and $0$ for transverse curves or curves outside $P_r$. This condition couples the hexahedral layer counts along different paths, including paths inside holes and along the exterior boundary. Surface interval compatibility alone does not ensure volume sweep compatibility \cite{Mitchell2023IIA}. Equation~\eqref{eq:sweep_path_intervals} applies to Sweep volumes; other schemes contribute their own applicable constraints.

For fixed meshing schemes and logical classifications, the surface and volume constraints are collected into a global integer feasibility problem \cite{Mitchell2023IIA}:
\begin{equation}
\begin{aligned}
\min_{\mathbf x}\quad
&f(\mathbf x,\mathbf g)\\
\text{subject to}\quad
&A\mathbf x=\mathbf b,\quad \mathbf x\in\mathbb Z^m,\\
&\boldsymbol\ell\leq\mathbf x\leq\mathbf u.
\end{aligned}
\label{eq:global_interval_feasibility}
\end{equation}
The vector $\mathbf x$ contains curve interval counts and any required auxiliary integers. The matrix $A$ collects surface balances, volume constraints such as Eq.~\eqref{eq:sweep_path_intervals}, applicable parity constraints, and explicitly fixed intervals; inequalities can be converted using bounded auxiliary variables. For example, a sum-even condition is encoded by $\sum_e n_e-2k=0$ with integer $k$. The bounds enforce admissible intervals, while $\mathbf g$ supplies the target-size goals. These goals are adjusted within the feasible set; the compatibility constraints remain hard requirements. Feasible interval assignment does not by itself certify successful geometric mesh generation or element quality.

For Map-compatible leaves, structured discretization follows the logical $i$-$j$-$k$ directions; Submap-compatible leaves use mappable logical subregions; and Sweep-compatible leaves propagate a quadrilateral source mesh toward a topologically compatible target through the linking surfaces. Final mesh generation and quality evaluation are performed downstream, without an additional learned mesh-quality optimizer. The manually constructed SALOME 9.16.0 verification example in Fig.~\ref{fig:global_conformity} uses a common quadrilateral layout and consistent layer divisions, with interface nodes created once and reused by adjacent hexahedra \cite{SALOMEHexaBlock916}. The complete inspected interface contains 896 quadrilaterals and 1,008 distinct nodes; every interface quadrilateral has exactly two incident hexahedra with opposite outward face orientations. The complete mesh contains 27,392 hexahedra and 32,328 nodes, forms one face-connected component, and has no duplicate coincident-node groups at a tolerance of $10^{-9}$ model units. This example verifies shared-interface conformity and does not imply a general meshing guarantee from local compatibility alone.

\subsection{DecompRL Optimization}
\label{sec:method_training}

DecompRL is optimized with a Soft Actor-Critic (SAC) formulation adapted to the candidate-conditioned decomposition setting. We use a masked candidate policy with twin Q-functions and soft target updates, together with the multi-child Tree-Bellman target below; thus, the method is not PPO and is not intended to be presented as a new generic RL algorithm.
Relative to the controlled DRL substitutions used in our experiments, SAC differs from DDPG and TD3 by using a stochastic entropy-regularized policy rather than deterministic action selection with external exploration noise; TD3 additionally employs target-policy smoothing and delayed actor updates. PPO is on-policy and optimizes a clipped surrogate objective rather than using replay-based off-policy updates. These comparisons therefore hold the CAD environment and candidate interface fixed while changing the RL optimization mechanism.

The value function itself follows the standard RL definition; the distinction in our setting lies in the transition structure used by the Bellman target. Since one split can produce multiple valid child solids, a standard single-next-state backup is insufficient for our decomposition tree. We therefore store tree transitions in replay and aggregate future value over all valid children generated by one split. Let $\operatorname{Child}(B_t,a_t)$ denote the set of child solids produced by executing geometric action $a_t$ on the active solid $B_t$. The reward $r_t$ measures the immediate quality of the selected split at the current node, while $V_{\bar{\nu}}(s')$ is the standard state-value estimate applied to each resulting child state. Accordingly, the training target combines immediate reward with discounted aggregated future value over all generated children:

\begin{equation}
\begin{aligned}
y_t = r_t + \gamma(1-d_t)\,\operatorname{Agg}\!\Big(
&\left\{V_{\bar{\nu}}(s') \mid s'=\psi_{\text{state}}(B'),\right.\\
&\left.B'\in\operatorname{Child}(B_t,a_t)\right\}
\Big),
\end{aligned}
\label{eq:tree_target}
\end{equation}
where $d_t$ is the terminal indicator, $V_{\bar{\nu}}$ is the same target value network, and $\operatorname{Agg}(\cdot)$ is an aggregation operator over child values. In our default setting, $\operatorname{Agg}(\cdot)$ is the sum operator, so a split is evaluated by the combined utility of all generated children rather than by only one favorable branch.

\subsubsection{Training Algorithm}
Algorithm~\ref{alg:tree_sac} summarizes the training loop used for the Tree-Bellman optimization described above: state encoding from the B-Rep, scoring of explicit candidates, masked selection, geometric reward composition, depth discounting, tree-structured target construction, and updates for the reported DRL method.
\begin{algorithm}[t]
\footnotesize
\caption{Training loop of DecompRL.}
\label{alg:tree_sac}
\KwIn{CAD model $M$, environment $\texttt{Env}$, replay buffer $\mathcal{D}$}
\KwInit{Actor $\pi_\theta$, critics $Q_{\phi_1},Q_{\phi_2}$, value network $V_\nu$, and target value network $V_{\bar{\nu}}$}
\For{each episode}{
  Reset the environment on $M$ to obtain the initial observation\;
  \While{not $\texttt{done}$}{
    Extract policy observation $o_t=(s_t,X_t,m_t)$ from the current active solid\;
    Select a feasible candidate index from the masked policy:
    $i_t \sim \pi_\theta(\cdot \mid o_t)$\;
    Set $a_t \leftarrow c_t^{i_t}$ and execute the corresponding split\;
    Verify generated child solids and obtain the child set together with runtime info\;
    Compute geometric reward components
    $q_{V,t},q_{E,t},q_{G,t},q_{L,t},q_{M,t}$ and
    $r_t^{\text{geo}}=\sum_k \bar{w}_k q_{k,t}$\;
    Set the base reward:
    $r_t^{\text{base}}\leftarrow -1.0$ if the split is invalid or yields empty children; otherwise $r_t^{\text{base}}\leftarrow r_t^{\text{geo}}$\;
    Apply runtime shaping:
    $r_t\leftarrow \lambda_d^{h_t}r_t^{\text{base}}+\mathbb{I}_{\text{stop},t}b_{\text{stop}}$\;
    Build tree-structured target:
    $v_t\leftarrow \operatorname{Agg}\!\left(\left\{V_{\bar{\nu}}(s')\mid s'=\psi_{\text{state}}(B'),\; B'\in \texttt{children}\right\}\right)$\;
    $y_t\leftarrow r_t+\gamma(1-d_t)v_t$\;
    Push the transition and child solids to $\mathcal{D}$\;
    Update $Q_{\phi_1},Q_{\phi_2},\pi_\theta,V_\nu$ from a minibatch in $\mathcal{D}$\;
    Soft update target value network $V_{\bar{\nu}}$\;
    Set the next frontier observation as input\;
  }
}
\end{algorithm}
The key difference from a standard one-step MDP backup is the multi-child target: a single split can produce multiple child solids, so the value target aggregates over all resulting children rather than a single next state.

\section{Experiments}
\label{sec:experiments}

We organize the experiments around four questions: Can DecompRL successfully decompose and mesh representative challenging CAD models with interacting features (RQ1)? Does it improve mesh quality and decomposition success rate over baselines under a unified evaluation setup across datasets (RQ2)? Are the proposed architectural components and reward terms reasonable for improving decomposition and meshing performance (RQ3)? And how efficient is it compared with the frame-field pipeline on aligned samples (RQ4)?

\subsection{Experiment Setting}
\label{sec:exp_setting}

\textbf{Datasets.}
Each training run comprises 5,000 episodes. At the start of each episode, one source CAD model is selected from the training pool of 1,400 models. For validation, we use four fixed sets of 200 CAD models each, one for each dataset (ABC~\cite{koch2019abc}, Fusion360~\cite{willis2020fusion}, NIST~\cite{Lipman2015NISTCAD}, and DeepCAD~\cite{DeepCAD2021}). Unless otherwise specified, ablation experiments are conducted on ABC.

\textbf{Implementation Details and Compared Baselines.}
We compare the reported DecompRL instantiation with PPO~\cite{Schulman2017ProximalPO}, DDPG~\cite{Lillicrap2015ContinuousCW}, and TD3~\cite{fujimoto2018addressing} as controlled \textbf{algorithm substitutions within the same CAD decomposition task}, rather than as application-independent baselines. These learned methods use the same CAD environment and geometric backend, solid-specific candidate-generation and feasibility-filtering pipeline, state/candidate observation interface, reward definition, split execution interface, hexahedral-meshing success criterion, and benchmark protocol. Their differences are therefore confined to the RL optimization scheme and algorithm-specific training machinery. We additionally compare with the prior decomposition method of Zhang et al.~\cite{Zhang2025Block}, which is evaluated with its original action families, and include the field-based LMFF method as the hexahedral-meshing baseline for runtime comparison~\cite{liu2023locally}. Unless otherwise specified, all experiments use the CadQuery/OCP geometric backend and a single NVIDIA RTX 3090 GPU, and all learned DRL methods are optimized with Adam.

\paragraph{Training configuration}
Unless otherwise noted, all learned DRL methods use the same hidden dimension. Table~\ref{tab:hparams} reports the optimization settings and trainable parameter counts for each method. Off-policy methods additionally use replay buffers and soft target-network updates, while PPO uses clipped policy optimization with multiple epochs per rollout. DDPG and TD3 further use scheduled exploration noise.
For the controlled comparison, DDPG and TD3 are adapted to the same finite masked candidate-selection interface rather than being used to output continuous geometric split parameters. Their actors score the same feasible candidates generated by the shared geometry pipeline; PPO likewise selects from the same masked candidate set. The comparison therefore isolates the RL optimization mechanism while keeping the geometric action interface fixed.

\begin{table}[t]
\centering
\small
\caption{Training configuration and parameter counts of the compared methods.}
\label{tab:hparams}
\resizebox{\columnwidth}{!}{%
\begin{tabular}{lcccc}
\toprule
 & DecompRL & DDPG & TD3 & PPO \\
\midrule
Hidden dimension        & 256 & 256 & 256 & 256 \\
Learning rate           & $1\times10^{-4}$ & $1\times10^{-4}$ & $1\times10^{-4}$ & $1\times10^{-4}$ \\
Batch size              & 128 & 128 & 128 & 128 \\
Target update $\tau$    & 0.005 & 0.005 & 0.005 & -- \\
Grad.\ norm clip        & 1.0 & 1.0 & 1.0 & 1.0 \\
Discount factor $\gamma$& 0.95 & 0.95 & 0.95 & 0.95 \\
Replay buffer           & \checkmark & \checkmark & \checkmark & -- \\
\# Parameters           & 838{,}069 & 828{,}419 & 975{,}300 & 881{,}538 \\
\bottomrule
\end{tabular}
}
\end{table}

\paragraph{Mesh size and shared settings}
For each input CAD model, we compute the target element size once from the original geometry before decomposition:
\begin{equation}
e_{\mathrm{size}}=\sqrt[3]{V_{\max}}\left(0.03+0.00045m_d^{3.1}\right),\qquad m_d=5.
\label{eq:target_mesh_size}
\end{equation}
Here, $V_{\max}$ is the largest solid volume in the original input model. The reference volume and target size are fixed for that input and reused across all methods in the benchmark Jacobian comparison, without recomputation from the child volumes after decomposition. Integer subdivisions may be adjusted to satisfy shared-interface and mapping/sweeping constraints; the common target size therefore does not require identical realized element counts. No additional mesh smoothing is applied in the main experiments.

\textbf{Metrics.} Success Rate (SR, reported as a percentage of evaluated models), the percentage of elements with scaled Jacobian $>0.8$ (written as J$>$0.8 in tables), minimum scaled Jacobian ($\min J_s$), and Mean Reward (written as R in tables). SR and the Jacobian metrics are the primary metrics, while Mean Reward is reported only as a supporting optimization signal. A model is counted as successful only if hexahedral meshing completes successfully. For the efficiency analysis, we report the runtime ratio for each sample, defined as $\mathrm{Speedup}=t_{\mathrm{LMFF}}/t_{\mathrm{DecompRL}}$. Here, $t_{\mathrm{LMFF}}$ covers the complete LMFF procedure used in our experiments, whereas $t_{\mathrm{DecompRL}}$ covers CAD decomposition only and excludes subsequent hexahedral mesh generation. The quantity labeled Speedup therefore denotes this stage-specific runtime ratio, rather than an end-to-end meshing speedup.
Mesh-quality statistics include only cases that successfully generate a complete globally conforming hexahedral mesh. For each successful case, we compute the proportion of elements with $J_s>0.8$ and the minimum scaled Jacobian, then average each metric with equal weight across successful cases. Thus, the reported $\min J_s$ is the average of per-case minima, and the reported J$>$0.8 value is the average of per-case proportions, rather than an element-count-weighted pooled statistic. Failed cases are excluded from quality averages without being assigned zero Jacobian values, but remain in the SR denominator.

\subsection{Case Study (RQ1)}

\label{sec:case_study}

\begin{figure}[!tbp]
\centering
\includegraphics[width=\linewidth]{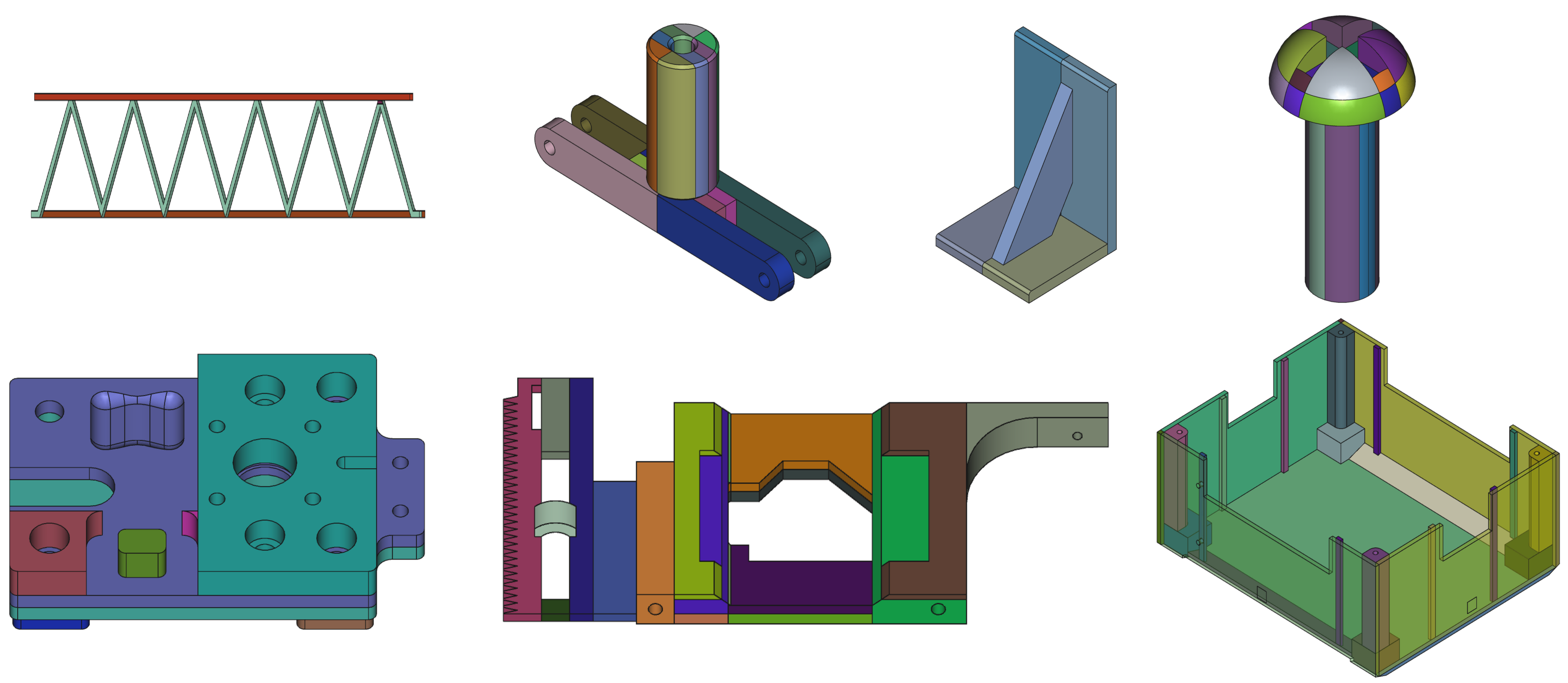}
\caption{Representative decomposition results produced by DecompRL. Colors denote child solids generated by the selected splits.}
\Description{Color coded CAD model decomposition results produced by DecompRL, in which each color corresponds to one child solid generated by the learned policy during experimental evaluation.}
\label{fig:qualitative_colored_decomp}
\end{figure}

\begin{figure*}[!t]
\centering
\captionsetup{font=footnotesize,skip=2pt}
\includegraphics[width=0.90\textwidth]{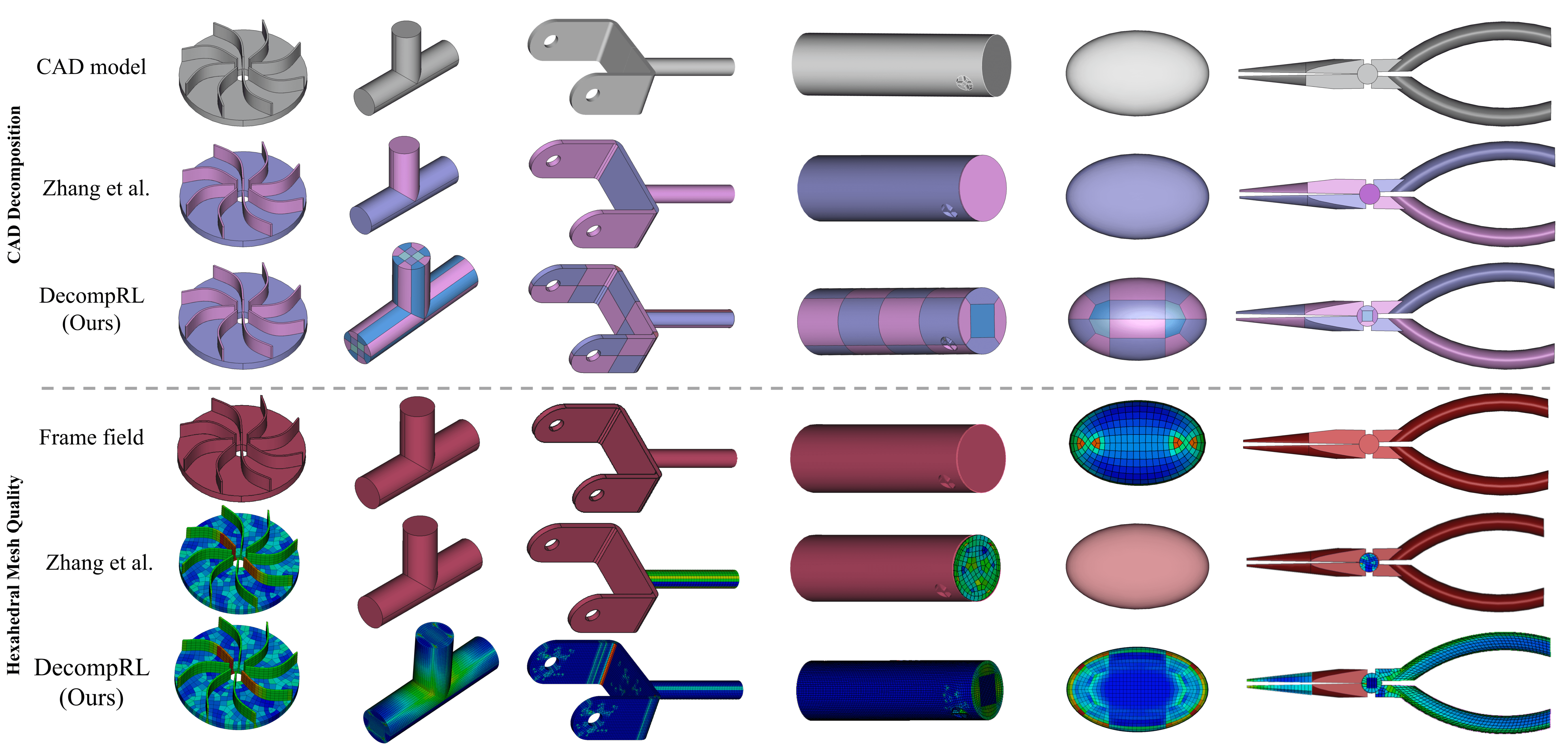}
\caption{Qualitative comparison of decomposition and hexahedral meshing on six representative CAD models. The upper half shows decomposition results, where different colors denote different sub-blocks. The lower half shows hexahedral meshing results colored by scaled Jacobian, where blue indicates better mesh quality and red indicates failure to generate hexahedral meshes.}
\Description{A six-case qualitative comparison figure. The upper half shows decomposition results for six CAD models, comparing Zhang et al. and DecompRL, with different colors indicating different sub-blocks. The lower half shows hexahedral meshing results from the LMFF frame-field pipeline, Zhang et al., and DecompRL, where blue indicates better mesh quality and red indicates failure to generate hexahedral meshes.}
\label{fig:qualitative_failmarks}
\end{figure*}

Fig.~\ref{fig:qualitative_colored_decomp} visualizes representative successful decomposition outputs produced by DecompRL across diverse CAD models, with each color denoting one child solid generated by the selected split sequence. Fig.~\ref{fig:qualitative_failmarks} then compares decomposition and hexahedral meshing results on six representative CAD models: its upper half shows the decompositions, where different colors denote sub-blocks, and its lower half shows the corresponding hexahedral meshes, colored by scaled Jacobian, where blue indicates better mesh quality and red indicates failure to generate hexahedral meshes.

The six comparison cases were deliberately selected to cover different combinations of splitting actions and to illustrate the resulting partitions and meshes. Dataset-level performance is assessed separately on the complete evaluation sets.

In the six-case comparison, DecompRL produces more refined and better aligned block partitions, especially for interacting features, and the resulting decompositions yield more complete, higher-quality hexahedral meshes. In the (2) T-junction case, it separates the orthogonal cylindrical branches and introduces an O-grid split to resolve the internal conflict. In the (4) cylindrical case, it isolates the wall block containing the hole from the regular cylindrical portion; in the (5) ellipsoidal case, it generates an O-grid-style decomposition suited to the curved outer shape; and in the (6) plier case, it separates the handle block from the remainder. These decompositions better resolve feature interactions and lead to clearer improvements in mesh quality in cases (2)--(5), with fewer red regions where hexahedral meshes cannot be generated. Although the less regular upper geometry leaves some red regions in case (6), DecompRL still produces the most complete and highest-quality mesh among the compared methods.

\begin{figure}[!t]
\centering
\includegraphics[width=\columnwidth]{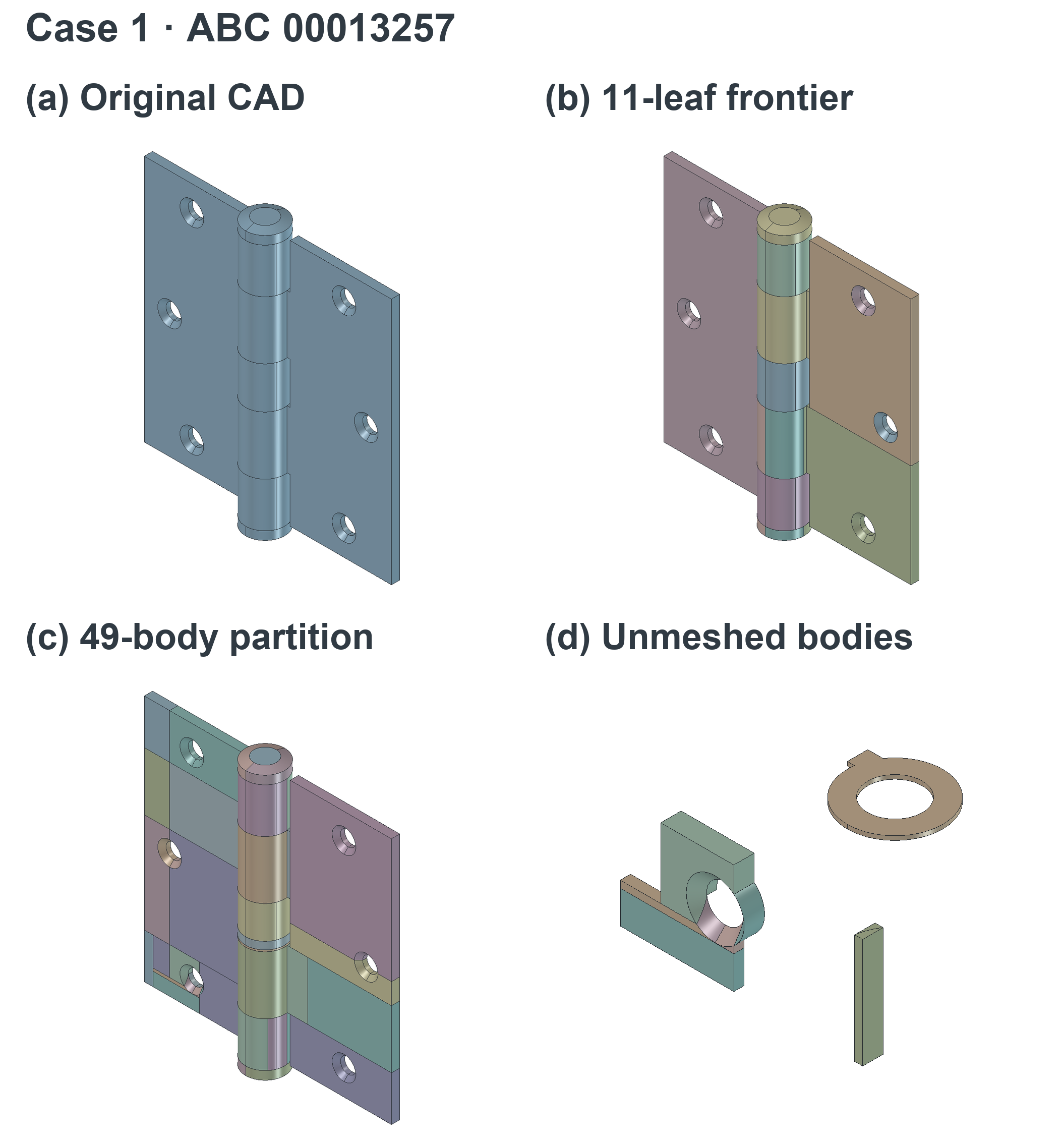}
\caption{Case 1 (ABC 00013257). (a) Original CAD. (b) Original 11-leaf frontier after six accepted actions. (c) Later 49-body diagnostic partition, shown as geometry. (d) Nine unmeshed bodies. The follow-up mesh remained incomplete. Colors distinguish bodies, not quality classes.}
\label{fig:failure_00013257}
\end{figure}

\begin{figure}[!t]
\centering
\includegraphics[width=\columnwidth]{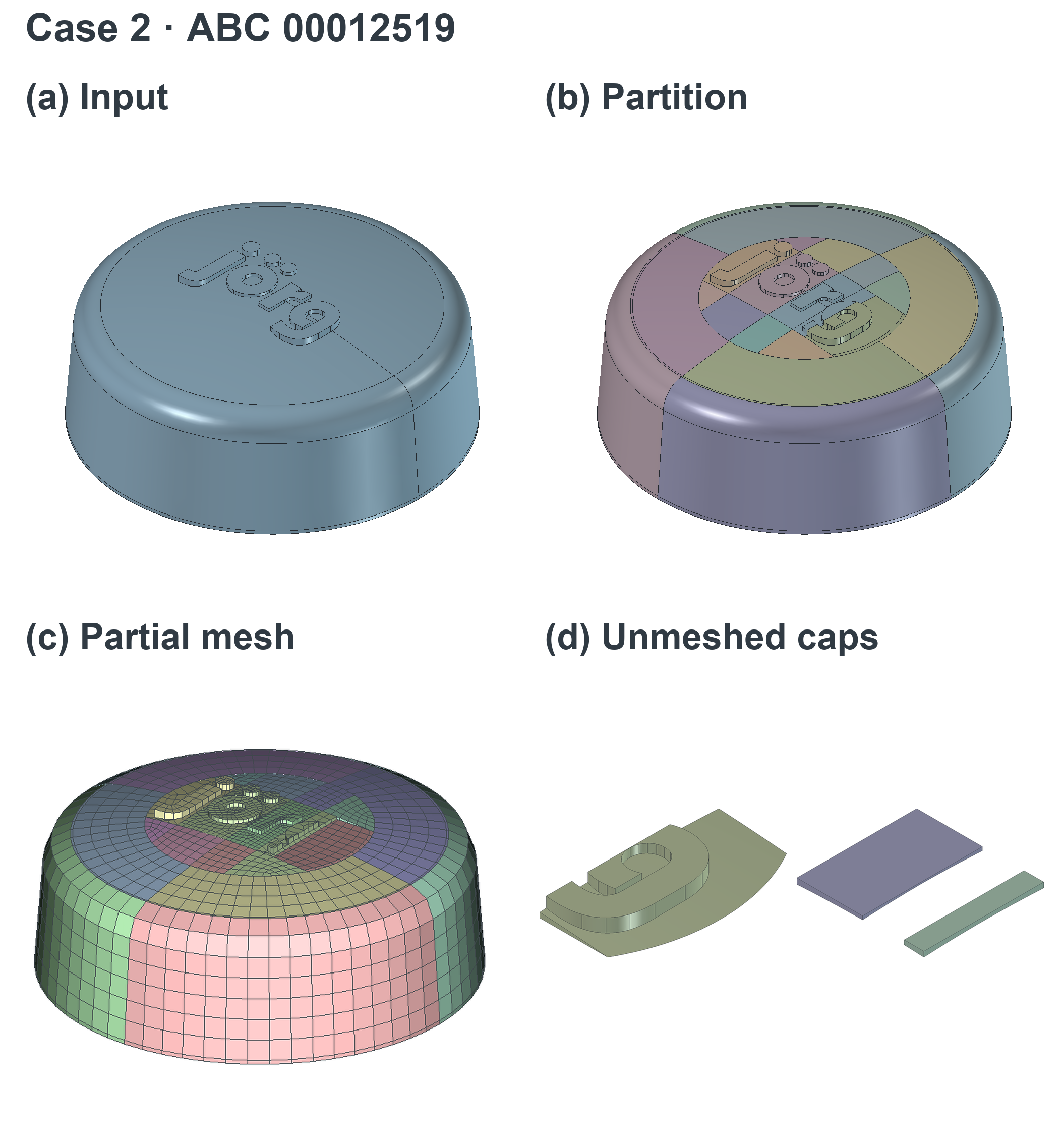}
\caption{Case 2 (ABC 00012519). (a) Original positive-volume source component. (b) Later 28-body diagnostic partition. (c) Partial mesh on 25 bodies. (d) The three unmeshed feature caps, shown together. The newly introduced central interfaces remain unverified. Colors distinguish bodies, not quality classes.}
\label{fig:failure_00012519}
\end{figure}

\textbf{Failure cases.} Two ABC models illustrate different causes of failure. In Case~1 (Fig.~\ref{fig:failure_00013257}), the rollout stopped with unresolved leaves. In Case~2 (Fig.~\ref{fig:failure_00012519}), Boolean validation rejected the proposed cuts. The follow-up diagnostics are reported separately from the original policy outcomes.

\textbf{Coupled residual meshing (ABC 00013257).} The perforated-plate and curved-sleeve model stopped after six accepted actions with 11 frontier leaves, four unresolved. Later case-specific processing produced 49 bodies, but only 40 contained hexahedra (10,853 elements), with nine elements below a scaled Jacobian of 0.2, including one inverted element; 41 interfaces remained incomplete. Some residual blocks could be meshed separately but not in the assembly, exposing coupled sweep and interval constraints.

\textbf{Geometric-validation failure (ABC 00012519).} For the raised-lettering model, all three attempted actions were rejected. Each cut produced two positive-volume children, but the subsequent Boolean validation failed with a vertex-mapping error. Later feature-oriented repairs meshed 25 of 28 bodies with 2,608 hexahedra and a minimum scaled Jacobian of 0.2316 over the meshed subset. Three feature-cap bodies remained unmeshed and new central interfaces were unverified, so the model remained unsuccessful.

\subsection{Overall Performance (RQ2)}
\label{sec:overall_performance}

\textbf{Main result comparison.} Table~\ref{tab:abc_benchmark} reports the endpoint comparison across datasets. DecompRL ranks first in mean SR on ABC, Fusion360, and NIST, with the largest margin observed on Fusion360. It also attains the best J$>$0.8 results on two of the four datasets and remains competitive in $\min J_s$, even though no single method dominates this metric uniformly. Overall, DecompRL achieves higher SR on three of the four evaluated datasets, with the clearest improvement observed on Fusion360, while maintaining competitive mesh-quality metrics. The method of Zhang et al.~\cite{Zhang2025Block} performs reasonably on ABC but degrades on the other datasets, likely due to its more limited action space. The DRL baselines remain competitive in selected metrics but generally trail DecompRL, supporting the value of its architectural and reward design.

\textbf{Statistical comparison.} We assess cross-dataset SR differences between DecompRL and PPO, DDPG, and TD3 using two-sided Welch $t$-tests computed from the reported, rounded means and sample standard deviations over three random-seed runs per method ($n=3$). Holm correction is applied across all 12 comparisons (three RL baselines across four datasets). These are comparisons of seed-level SR summaries, rather than paired tests of individual model outcomes. Table~\ref{tab:sr_significance} reports the comparison against the RL baseline with the highest mean SR on each dataset.

On Fusion360, DecompRL exceeds PPO by 12.6 percentage points, and this difference remains significant at the 0.05 level after correction ($p=0.000677$). The smaller numerical gains on ABC and NIST are not significant after correction. On DeepCAD, DecompRL is 1.3 percentage points below DDPG, also without a significant difference after correction. A non-significant result does not establish equivalence between methods.

\begin{table}[!t]
\centering
\captionsetup{font=footnotesize,skip=2pt}
\caption{SR comparisons against the RL baseline with the highest mean SR on each dataset. Differences are DecompRL minus the baseline, in percentage points. Holm correction covers all 12 RL-baseline comparisons, not only the four shown.}
\label{tab:sr_significance}
\resizebox{\linewidth}{!}{%
\begin{tabular}{llrrr}
\toprule
\textbf{Dataset} & \textbf{Baseline} & \textbf{SR difference} & \textbf{Raw $p$} & \textbf{Holm-adjusted $p$} \\
\midrule
ABC & PPO & $+1.6$ & 0.4521 & 0.9043 \\
Fusion360 & PPO & $+12.6$ & $6.77\times10^{-5}$ & $6.77\times10^{-4}$ \\
NIST & PPO & $+2.1$ & 0.0261 & 0.1307 \\
DeepCAD & DDPG & $-1.3$ & 0.1129 & 0.4517 \\
\bottomrule
\end{tabular}%
}
\end{table}

\begin{table*}[!tbp]
  \centering
\caption{Quantitative comparison across methods and datasets under the unified evaluation setup. Values are reported as mean$\pm$std over seeds 42, 43, and 44; higher is better for all reported metrics.}
  \label{tab:abc_benchmark}
  \scriptsize
  \setlength{\tabcolsep}{2.5pt}
  \resizebox{\textwidth}{!}{%
  \begin{tabular}{l*{16}{c}}
  \toprule
  \textbf{Method}
  & \multicolumn{4}{c}{\textbf{ABC}}
  & \multicolumn{4}{c}{\textbf{Fusion360}}
  & \multicolumn{4}{c}{\textbf{NIST}}
  & \multicolumn{4}{c}{\textbf{DeepCAD}} \\
  \cmidrule(lr){2-5}\cmidrule(lr){6-9}\cmidrule(lr){10-13}\cmidrule(lr){14-17}
  & \textbf{SR (\%)} & \textbf{J$>$0.8 (\%)} & \textbf{$\min J_s$} & \textbf{R}
  & \textbf{SR (\%)} & \textbf{J$>$0.8 (\%)} & \textbf{$\min J_s$} & \textbf{R}
  & \textbf{SR (\%)} & \textbf{J$>$0.8 (\%)} & \textbf{$\min J_s$} & \textbf{R}
  & \textbf{SR (\%)} & \textbf{J$>$0.8 (\%)} & \textbf{$\min J_s$} & \textbf{R} \\
  \midrule
  PPO & 39.0 $\pm$ 1.3 & 89.0 $\pm$ 1.1 & \textbf{0.5} $\pm$ 0.0 & 1.4 $\pm$ 0.1 & 47.0 $\pm$ 0.9 & 85.8 $\pm$ 0.4 & 0.5 $\pm$ 0.1 & 3.4 $\pm$ 0.2 & 37.5 $\pm$ 0.6 & 80.1 $\pm$ 0.1 & 0.3 $\pm$ 0.1 & 2.7 $\pm$ 0.5 & 75.8 $\pm$ 0.1 & 85.2 $\pm$ 1.0 & 0.6 $\pm$ 0.0 & 3.1 $\pm$ 0.2 \\
  DDPG & 32.0 $\pm$ 1.8 & 86.7 $\pm$ 1.7 & 0.4 $\pm$ 0.1 & 1.9 $\pm$ 0.1 & 44.0 $\pm$ 0.7 & \textbf{89.2} $\pm$ 0.4 & \textbf{0.6} $\pm$ 0.1 & 4.1 $\pm$ 1.0 & 34.3 $\pm$ 0.3 & 71.7 $\pm$ 0.6 & 0.3 $\pm$ 0.0 & 3.3 $\pm$ 0.8 & \textbf{77.3} $\pm$ 0.5 & \textbf{89.0} $\pm$ 0.1 & 0.5 $\pm$ 0.1 & \textbf{3.7} $\pm$ 0.3 \\
  TD3 & 36.5 $\pm$ 2.2 & 90.1 $\pm$ 2.0 & 0.4 $\pm$ 0.1 & 1.6 $\pm$ 0.3 & 43.5 $\pm$ 0.7 & 88.9 $\pm$ 0.0 & 0.5 $\pm$ 0.1 & 3.3 $\pm$ 0.5 & 32.5 $\pm$ 0.2 & 83.5 $\pm$ 0.7 & \textbf{0.4} $\pm$ 0.0 & 2.5 $\pm$ 1.0 & 71.2 $\pm$ 0.1 & 88.4 $\pm$ 0.6 & \textbf{0.8} $\pm$ 0.0 & 3.2 $\pm$ 0.7 \\
  Zhang et al.~\cite{Zhang2025Block} & 25.0 $\pm$ 2.6 & 84.9 $\pm$ 2.4 & 0.3 $\pm$ 0.0 & 1.2 $\pm$ 0.0 & 34.0 $\pm$ 0.8 & 84.4 $\pm$ 0.3 & 0.4 $\pm$ 0.1 & 4.3 $\pm$ 0.2 & 27.5 $\pm$ 1.0 & 86.8 $\pm$ 0.5 & \textbf{0.4} $\pm$ 0.0 & 3.3 $\pm$ 0.9 & 65.6 $\pm$ 0.3 & 82.1 $\pm$ 0.5 & 0.6 $\pm$ 0.1 & 2.6 $\pm$ 0.2 \\
  DecompRL (ours) & \textbf{40.6} $\pm$ 2.9 & \textbf{91.6} $\pm$ 2.9 & \textbf{0.5} $\pm$ 0.1 & \textbf{3.2} $\pm$ 0.3 & \textbf{59.6} $\pm$ 0.7 & 88.2 $\pm$ 0.2 & \textbf{0.6} $\pm$ 0.1 & \textbf{5.2} $\pm$ 0.5 & \textbf{39.6} $\pm$ 0.0 & \textbf{87.5} $\pm$ 0.8 & \textbf{0.4} $\pm$ 0.0 & \textbf{3.9} $\pm$ 0.6 & 76.0 $\pm$ 0.9 & 88.7 $\pm$ 0.2 & \textbf{0.8} $\pm$ 0.1 & 3.5 $\pm$ 0.4 \\
  \bottomrule
  \end{tabular}
  }
\end{table*}

\begin{figure}[!t]
\centering

\begin{minipage}[t]{0.4\linewidth}
  \centering
  \includegraphics[width=\linewidth]{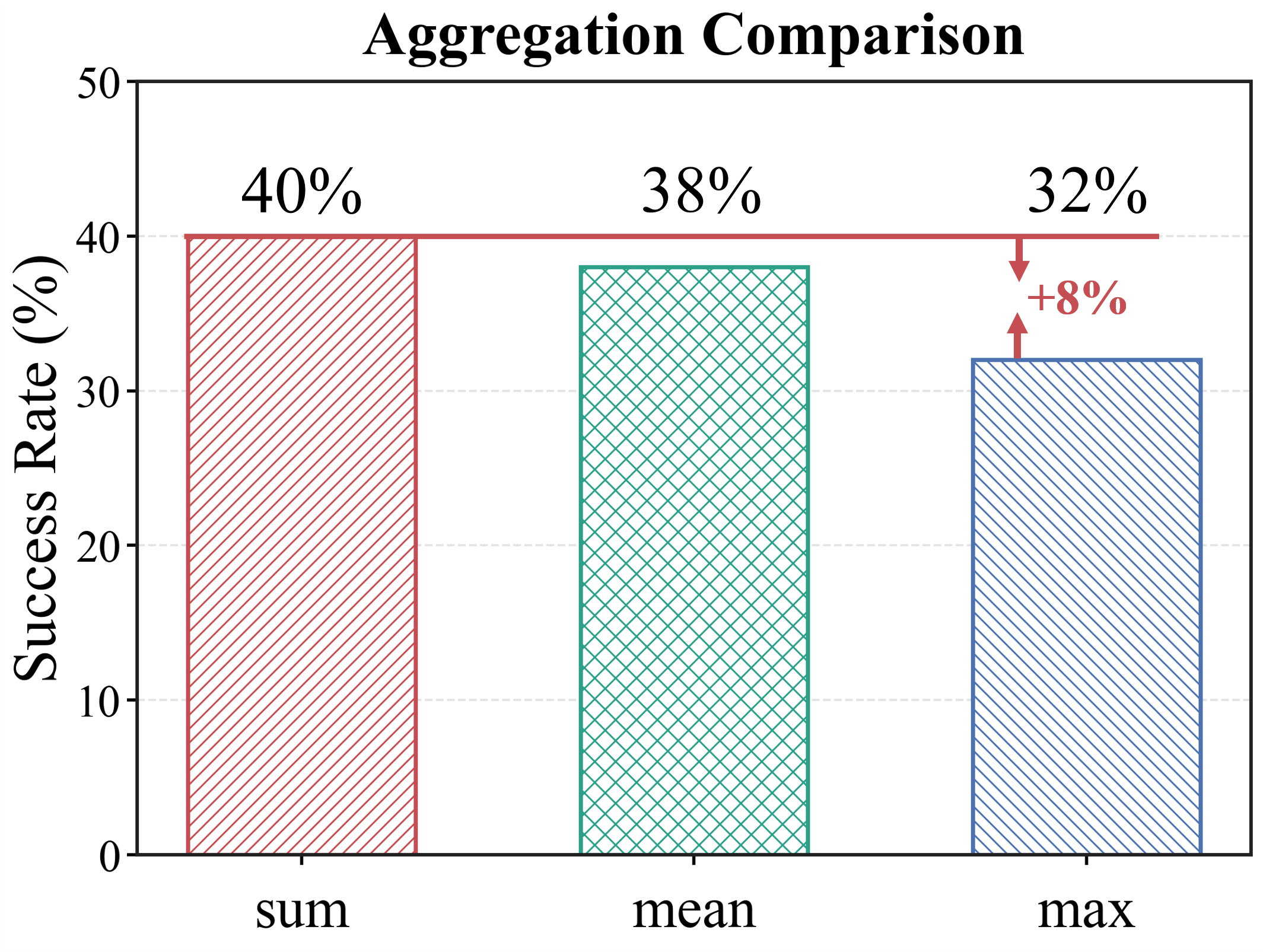}
  \\[-3pt]\scriptsize (a) Aggregation on ABC
\end{minipage}\hfill
\begin{minipage}[t]{0.6\linewidth}
  \centering
  \includegraphics[width=\linewidth]{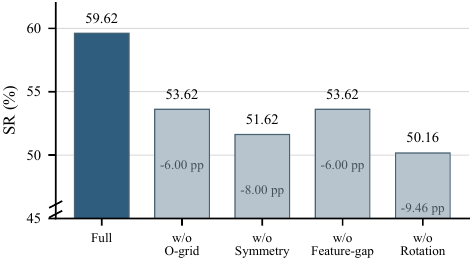}
  \\[-3pt]\scriptsize (b) Action-space ablation
\end{minipage}

\captionsetup{font=footnotesize,skip=2pt}
\caption{Aggregation and action-space ablations of DecompRL. (a) compares sum, mean, and max aggregation over child values on ABC; values are mean SR over seeds 42, 43, and 44. (b) reports action-space ablation on Fusion360 ($N=200$); bar labels give SR and in-bar labels show the SR change in percentage points relative to Full DecompRL. The y-axis is truncated below 45\% to resolve differences among high-SR variants.}
\Description{A two-part figure. The left panel compares sum, mean, and max child-value aggregators on ABC. The right panel is a five-bar Fusion360 success-rate chart for Full DecompRL and variants without O-grid, symmetry, feature-gap, or rotation actions.}
\label{fig:training_aggregation_pair}
\end{figure}

\subsubsection{Learning Progression}
\begin{figure}[!t]
\centering
\includegraphics[width=\linewidth]{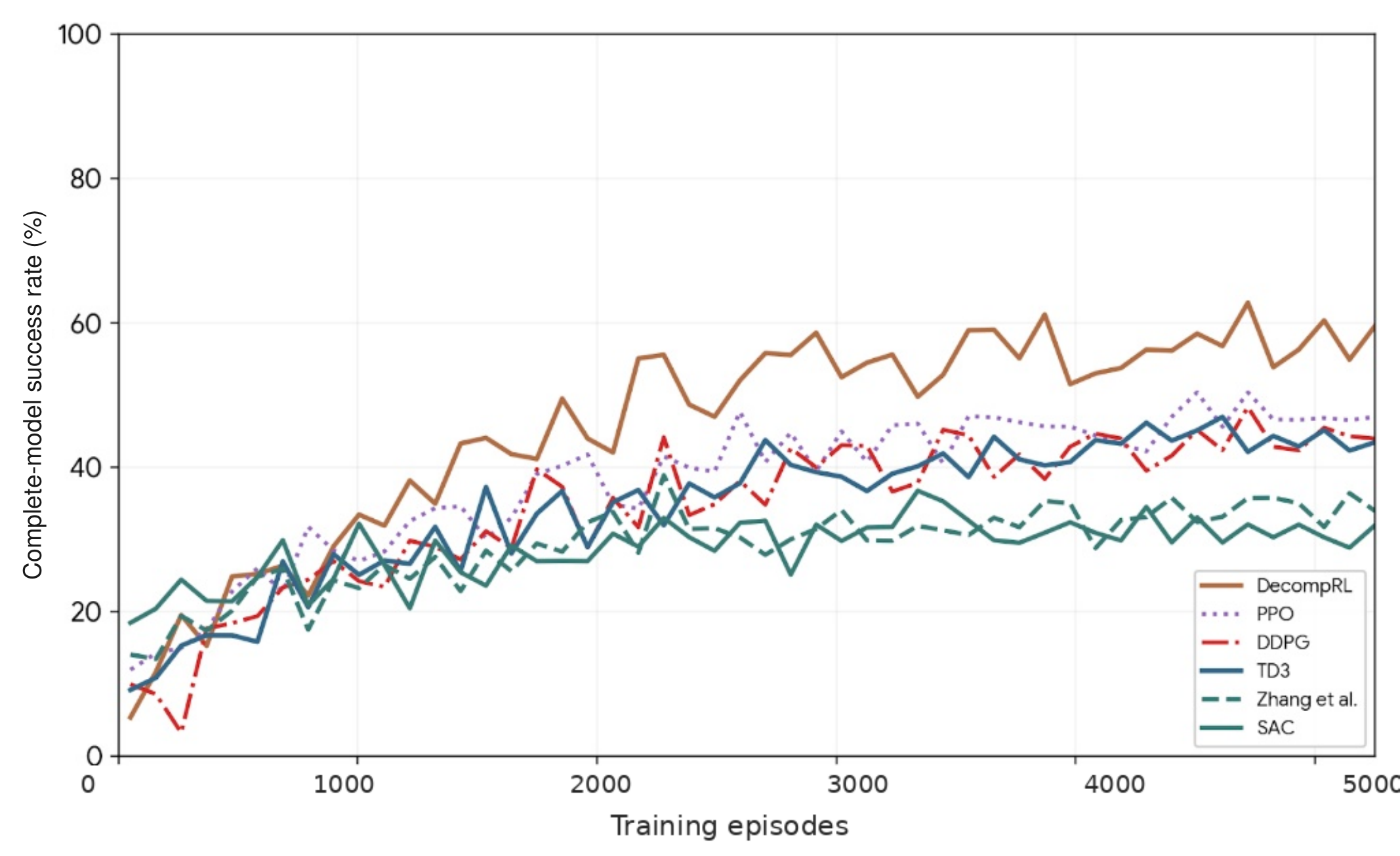}
\caption{Complete-model success rate during training on a fixed set of 200 Fusion360 CAD models. Each curve shows one run per method. A model counts as successful when hexahedral meshing completes for the entire model.}
\label{fig:fusion360_learning_progress}
\end{figure}

Figure~\ref{fig:fusion360_learning_progress} shows complete-model SR during training on the same fixed Fusion360 evaluation set. At each evaluated episode $t$, $\mathrm{SR}(t)=100N_{\mathrm{success}}(t)/N_{\mathrm{eval}}$, with a fixed denominator. A model counts as successful only if hexahedral meshing completes for the entire model, including all final leaves. DecompRL rises during early training and reaches approximately 60\% toward the end of the plotted period.

\begin{figure*}[!t]
\centering
\includegraphics[width=0.96\textwidth,trim=0 360bp 0 0,clip]{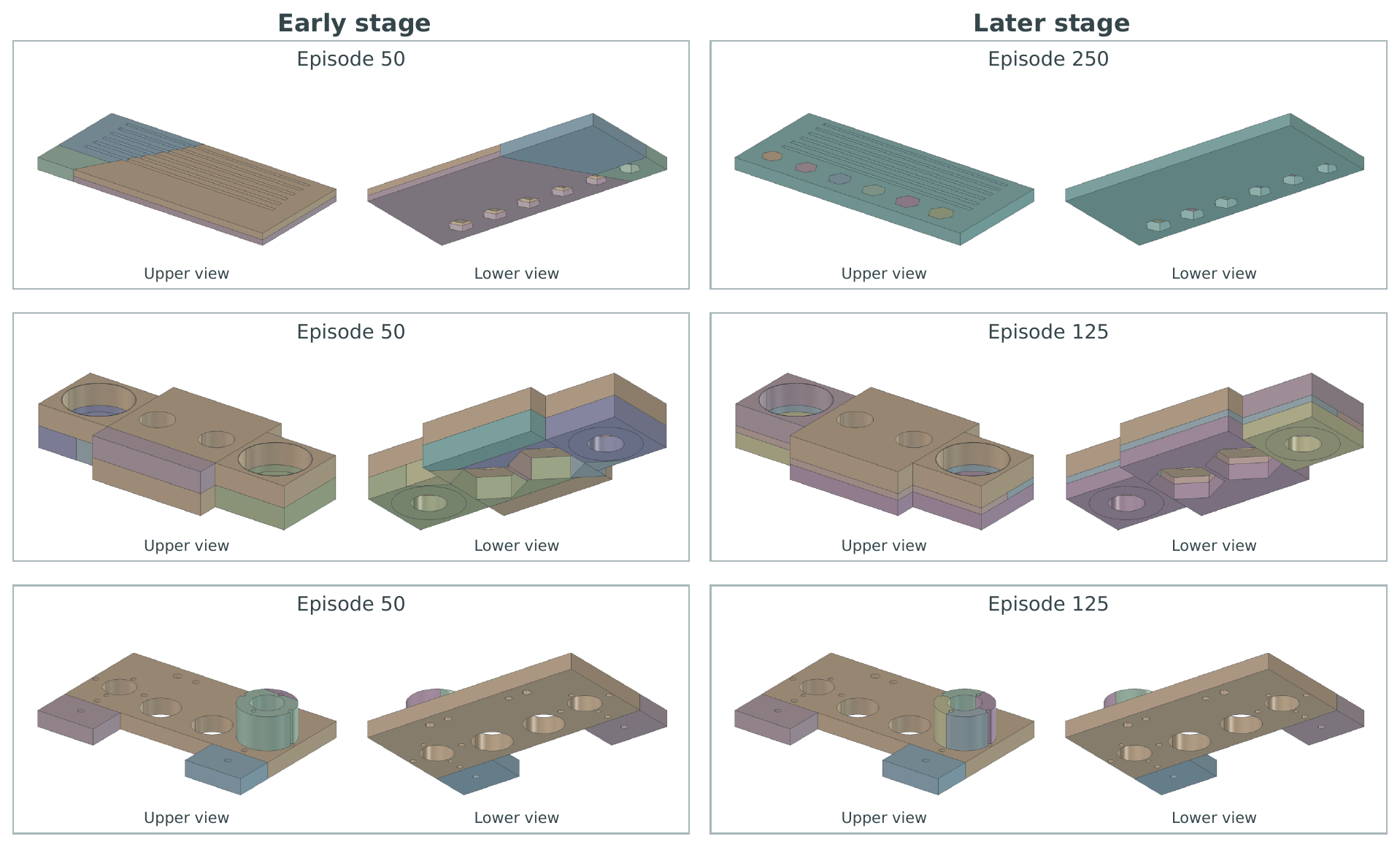}\\
\includegraphics[width=0.96\textwidth,trim=0 0 0 379bp,clip]{revision_figures/training_three_cases.pdf}
\caption{Archived decompositions from a separate historical training run, redrawn in FreeCAD. The top row shows ABC 00006119 and the bottom row shows ABC 00026545. The left and right boxes show the earlier and later episodes, respectively, with the episode number above each pair of views. Each pair shows the partition from above and from the opposite lower side. Colors distinguish solids within each view and do not track solids across episodes.}
\label{fig:training_decomposition_snapshots}
\end{figure*}

Figure~\ref{fig:training_decomposition_snapshots} complements this curve with archived decomposition outputs for two models. These images come from a historical training/evaluation run distinct from the 5,000-episode benchmark runs. Solid counts change from 4 to 7 for 00006119 (episodes 50 and 250) and from 7 to 5 for 00026545 (episodes 50 and 125). Because evaluation seeds and terminal-verification settings differ across the historical snapshots, they illustrate changes in partition layout rather than controlled failure-to-success transitions. Block count alone does not establish improved meshing quality.

\subsection{Ablation Study (RQ3)}
\label{sec:ablation}
Table~\ref{tab:ablation} summarizes two groups of ablations, covering architectural components and reward components, and reports their impact on SR and mesh quality. Each variant is trained with three random seeds, and the results are reported as mean$\pm$std.

\textbf{Main component ablations.}
Removing Tree Bellman backup degrades SR across all three seeds, supporting the role of multi-child aggregation in long-horizon decomposition. Disabling the advanced actions causes the largest drop, indicating that richer split primitives are directly relevant to hexahedral meshability. To isolate the learned candidate-scoring architecture from geometric pre-filtering, the early-fusion ablation keeps the same candidate generation, feasibility mask, and candidate set as the full model, but replaces the separate state/candidate encoders with a shared encoder over concatenated raw descriptors. Its lower SR therefore reflects the effect of the learned candidate-conditioned scoring architecture rather than removal of candidate pre-filtering.

Fig.~\ref{fig:training_aggregation_pair}(a) compares the aggregation operator over child values in Eq.~\ref{eq:tree_target} on ABC, reporting mean SR over seeds 42, 43, and 44. We observe a ranking of sum $>$ mean $>$ max. This ordering matches the decomposition objective, in which overall utility depends on jointly strong child branches rather than a single favorable branch. Mean dilutes cumulative gains across branches, whereas max overemphasizes a single optimistic branch.

\subsubsection{Comparison with Non-Learning Selectors}
To examine the contribution of learned action selection, we compare DecompRL with a deterministic Greedy selector and Random-valid, which samples uniformly from valid candidates (Table~\ref{tab:nonlearning_selectors}). Only the action-selection strategy changes: all methods share the candidate-generation and feasibility-masking rules, execution pipeline, action budget, and complete-model meshing-success criterion, and are evaluated on the same fixed 200 CAD models per dataset. DecompRL uses the main benchmark means across three seeds; Random-valid is evaluated with seeds 42, 43, and 44 and reported as mean$\pm$standard deviation.

\begin{table}[!t]
\centering
\captionsetup{font=footnotesize,skip=2pt}
\caption{Non-learning selector comparison. All SR values are percentages. The last column is the maximum Holm-adjusted McNemar $p$-value across three DecompRL seeds against Greedy, separately for each dataset.}
\label{tab:nonlearning_selectors}
\resizebox{\linewidth}{!}{%
\begin{tabular}{lrrrr}
\toprule
Dataset & DecompRL & Greedy & Random-valid & Max. adjusted $p$ \\
\midrule
ABC & 40.6 & 16.0 & $14.7\pm1.6$ & $<0.01$ \\
Fusion360 & 59.6 & 16.5 & $17.8\pm0.6$ & $<0.01$ \\
NIST & 39.6 & 11.5 & $14.8\pm2.0$ & $<0.01$ \\
DeepCAD & 76.0 & 26.5 & $20.2\pm1.4$ & $<0.001$ \\
\bottomrule
\end{tabular}%
}
\end{table}

For each dataset and each of the three DecompRL seeds, we perform a two-sided exact McNemar test against deterministic Greedy using discordant paired success/failure outcomes on the same 200 models. Holm correction is applied across all 12 tests (three seeds across four datasets). The table reports the largest adjusted $p$-value across the three seeds for each dataset. No statistical test against Random-valid is reported.

DecompRL improves SR over Greedy by 24.6, 43.1, 28.1, and 49.5 percentage points on ABC, Fusion360, NIST, and DeepCAD, respectively. Its mean SR also exceeds Random-valid on all four datasets; this latter comparison is descriptive. These results support a contribution from learned action selection beyond the two tested non-learning selectors under the shared geometric proposal and feasibility-filtering pipeline.

\subsubsection{Action-Space Ablation}
While Table~\ref{tab:ablation} removes all advanced actions jointly, we further ablate individual primitives to quantify the contribution of each action family. We conduct this study on a fixed Fusion360 subset ($N=200$). Fig.~\ref{fig:training_aggregation_pair}(b) shows that removing any single family lowers SR. Rotation and symmetry splits have the largest impact ($\Delta$SR $-9.46$ and $-8.00$ percentage points), consistent with their role in resolving sweep-direction conflicts in ring-like structures. O-grid and feature-gap splits each reduce SR by 6.00 percentage points, confirming their importance for circular features and closely spaced misaligned features. Overall, all advanced primitives contribute to the final performance.

\textbf{Reward component ablations.}
Disabling individual reward terms hurts either success rate or mesh quality in each ablation variant. The angle-based terms mainly affect quality consistency, so both face-corner-angle and dihedral-angle regularization remain important. Removing the internal face-boundary-loop term decreases the decomposition success rate, suggesting that this regularization helps discourage structurally unfavorable intermediate blocks, whereas removing the extreme-geometry term lowers the fraction of high-quality elements. Excluding the meshability term degrades both SR and hexahedral mesh quality, indicating that block-level meshability information provides useful guidance beyond local split validity.

\subsubsection{Reward Weight Sensitivity}
\begin{figure}[!t]
\centering
\captionsetup{font=footnotesize,skip=2pt}
\includegraphics[width=\linewidth,trim={0 0 190bp 0},clip]{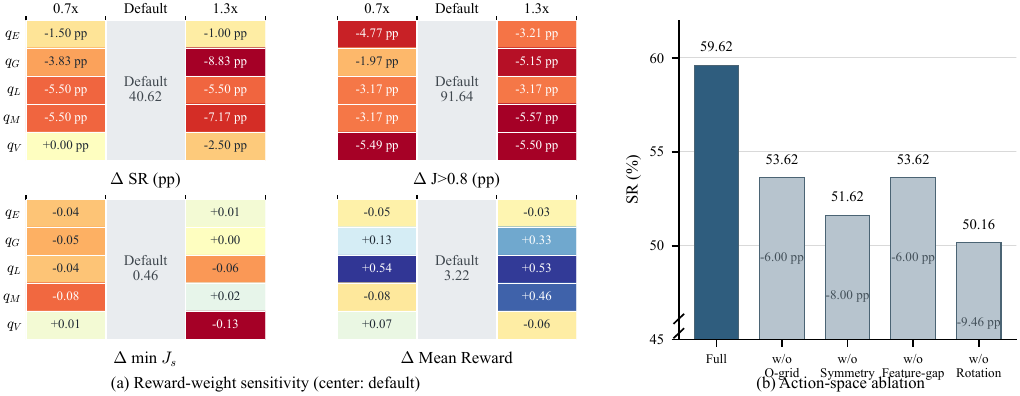}
\caption{Reward-weight sensitivity of DecompRL. Results are reported across SR, J$>$0.8, $\min J_s$, and Mean Reward. Rows correspond to reward terms; the center band gives the default value once, and the outer columns give signed changes induced by 0.7$\times$ and 1.3$\times$ scaling. Colors are normalized separately for each metric.}
\Description{Four heatmaps for reward-weight sensitivity across five reward terms. Each heatmap has a single default-value band and two outer columns for signed changes under lower and higher weight scales.}
\label{fig:rq3_sensitivity_analysis}
\end{figure}

Fig.~\ref{fig:rq3_sensitivity_analysis} shows reward-weight sensitivity across the four reported metrics: SR, J$>$0.8, $\min J_s$, and Mean Reward. Each heatmap organizes the results by reward term and scaling factor. The center band reports the default DecompRL value once, while the outer cells show the signed change induced by 0.7$\times$ and 1.3$\times$ scaling.

\begin{table}[!t]
\centering
\captionsetup{font=footnotesize,skip=2pt}
\caption{Ablation results for the architectural and reward components of DecompRL. Each variant is trained with three random seeds; values are mean$\pm$std. The Full Model row corresponds to DecompRL in Table~\ref{tab:abc_benchmark}; minor differences are due to rounding. Boldface marks the best value in each primary endpoint column.}
\label{tab:ablation}
\resizebox{\linewidth}{!}{%
\begin{tabular}{lcccc}
\toprule
\textbf{Variant} & \textbf{SR (\%)} & \textbf{J$>$0.8 (\%)} & \textbf{$\min J_s$} & \textbf{Mean Reward} \\
\midrule
\multicolumn{5}{l}{\textit{Main Component Ablations}} \\
Full Model & \textbf{40.62 $\pm$ 2.91} & \textbf{91.64 $\pm$ 2.88} & \textbf{0.46 $\pm$ 0.09} & 3.22 $\pm$ 0.30 \\
w/o Tree Bellman & 31.88 $\pm$ 2.65 & 87.00 $\pm$ 1.12 & 0.43 $\pm$ 0.17 & 3.04 $\pm$ 0.02 \\
w/o Advanced Actions & 30.00 $\pm$ 1.77 & 91.63 $\pm$ 0.86 & 0.36 $\pm$ 0.04 & 2.94 $\pm$ 0.13 \\
Early-fusion policy (same candidate pool) & 30.62 $\pm$ 4.42 & 86.81 $\pm$ 0.12 & 0.38 $\pm$ 0.04 & 2.93 $\pm$ 0.20 \\
\midrule
\multicolumn{5}{l}{\textit{Reward Component Ablations}} \\
w/o Face-Corner Angles ($q_V$) & 35.00 $\pm$ 3.54 & 87.51 $\pm$ 0.51 & 0.38 $\pm$ 0.01 & 2.92 $\pm$ 0.03 \\
w/o Dihedral Angles ($q_E$) & 33.75 $\pm$ 1.50 & 83.92 $\pm$ 4.93 & 0.35 $\pm$ 0.02 & 2.97 $\pm$ 0.07 \\
w/o Extreme Geometry ($q_G$) & 35.00 $\pm$ 5.30 & 85.71 $\pm$ 1.86 & 0.42 $\pm$ 0.04 & 3.04 $\pm$ 0.06 \\
w/o Internal Face-Boundary Loops ($q_L$) & 32.50 $\pm$ 0.07 & 89.45 $\pm$ 0.37 & 0.35 $\pm$ 0.05 & 2.95 $\pm$ 0.13 \\
w/o Meshability Term ($q_M$) & 31.40 $\pm$ 0.02 & 82.61 $\pm$ 1.24 & 0.34 $\pm$ 0.02 & 2.88 $\pm$ 0.08 \\
\bottomrule
\end{tabular}
}
\end{table}

\subsubsection{Depth-Discount Sensitivity}
We compare depth-discount configurations on the fixed Fusion360 test set under identical environment settings and a decision budget of 8 steps (Table~\ref{tab:depth_discount_sensitivity}). The default configuration uses $\lambda_d=0.90$ and $\gamma=0.95$. Here, $\lambda_d$ scales the immediate reward by tree depth, whereas $\gamma$ discounts future child-state values. SR measures complete-model conforming all-hexahedral meshing success within the budget. Average tree depth is the mean of the maximum decomposition-tree depth per evaluated model, and average steps is the mean number of executed decision steps. The high-aspect-ratio leaf percentage is computed over all accepted leaf blocks across the evaluated models, counting those whose oriented-bounding-box dimensions satisfy $\min(L)/\max(L)<0.05$.

\begin{table}[!t]
\centering
\captionsetup{font=footnotesize,skip=2pt}
\caption{Discount sensitivity on Fusion360 under an 8-step budget. Depth and steps are model-level means; high-AR denotes the aggregate percentage of accepted leaf blocks with high aspect ratios. The default configuration corresponds to the main-paper baseline.}
\label{tab:depth_discount_sensitivity}
\resizebox{\linewidth}{!}{%
\begin{tabular}{lrrrrrr}
\toprule
Configuration & $\lambda_d$ & $\gamma$ & SR (\%) & Depth & Steps & High-AR (\%) \\
\midrule
Stronger decay & 0.80 & 0.95 & 56.5 & 2.5 & 5.2 & 3.5 \\
Default & 0.90 & 0.95 & 59.6 & 2.8 & 5.77 & 4.1 \\
Lower Bellman discount & 0.90 & 0.90 & 58.2 & 2.7 & 5.5 & 3.8 \\
No depth decay & 1.00 & 0.95 & 47.2 & 3.8 & 7.2 & 14.6 \\
\bottomrule
\end{tabular}%
}
\end{table}

Removing depth decay reduces SR from 59.6\% to 47.2\%, a decrease of 12.4 percentage points, while increasing mean maximum tree depth from 2.8 to 3.8 and mean decision steps from 5.77 to 7.2. The high-aspect-ratio leaf percentage also increases from 4.1\% to 14.6\%. Stronger depth decay ($\lambda_d=0.80$) yields an SR of 56.5\%, while reducing $\gamma$ to 0.90 with the default depth discount yields 58.2\%. Among the tested configurations, the default achieves the highest SR, and configurations with depth decay produce shallower trees than the configuration without depth decay.

\subsection{Efficiency Analysis (RQ4)}
\label{sec:efficiency}

\begin{figure}[!t]
\centering
\begin{minipage}[t]{0.48\linewidth}
  \centering
  \includegraphics[width=\linewidth]{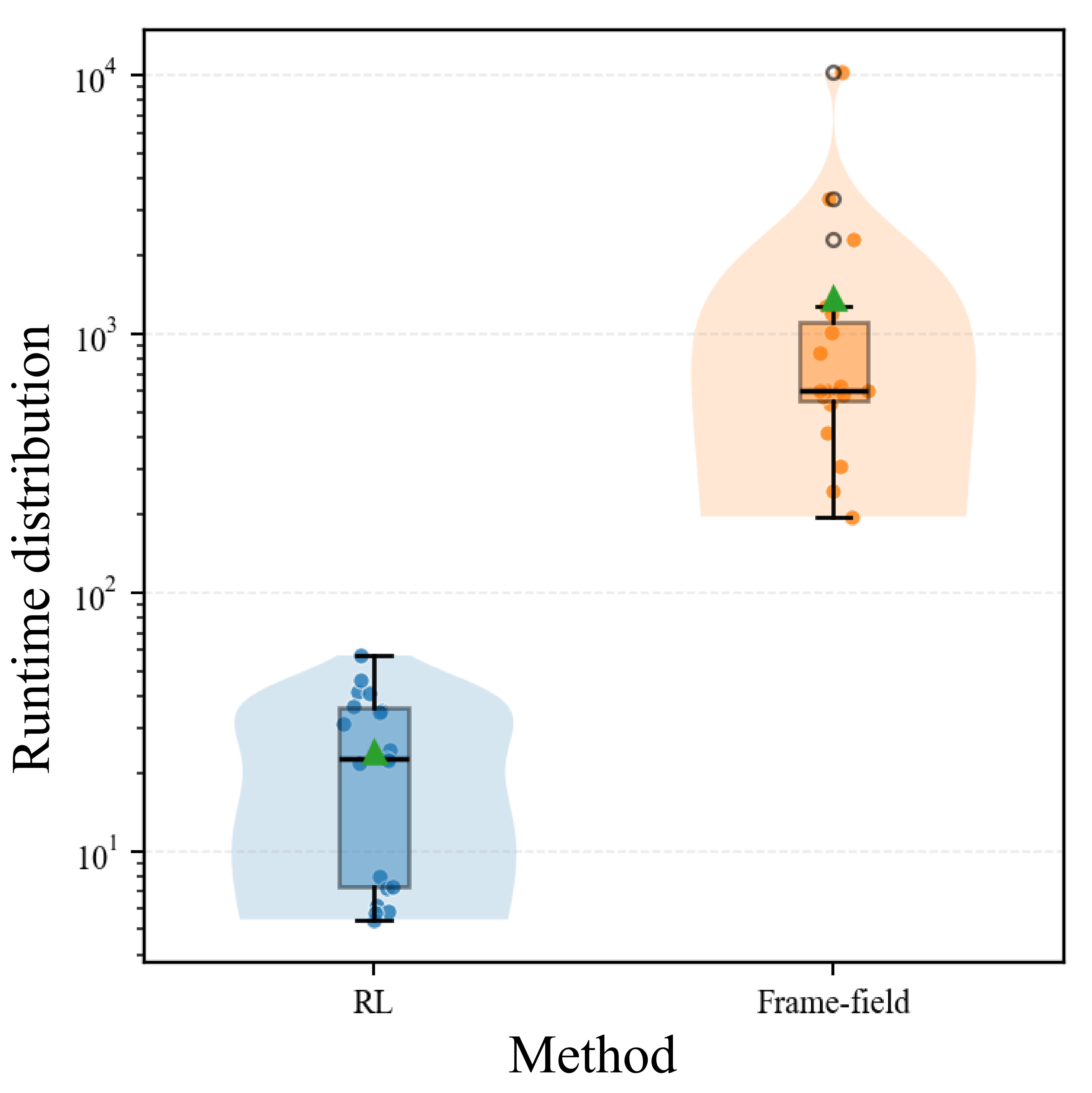}
  \\[-3pt]\scriptsize (a) Runtime distribution
\end{minipage}\hfill
\begin{minipage}[t]{0.48\linewidth}
  \centering
  \includegraphics[width=\linewidth]{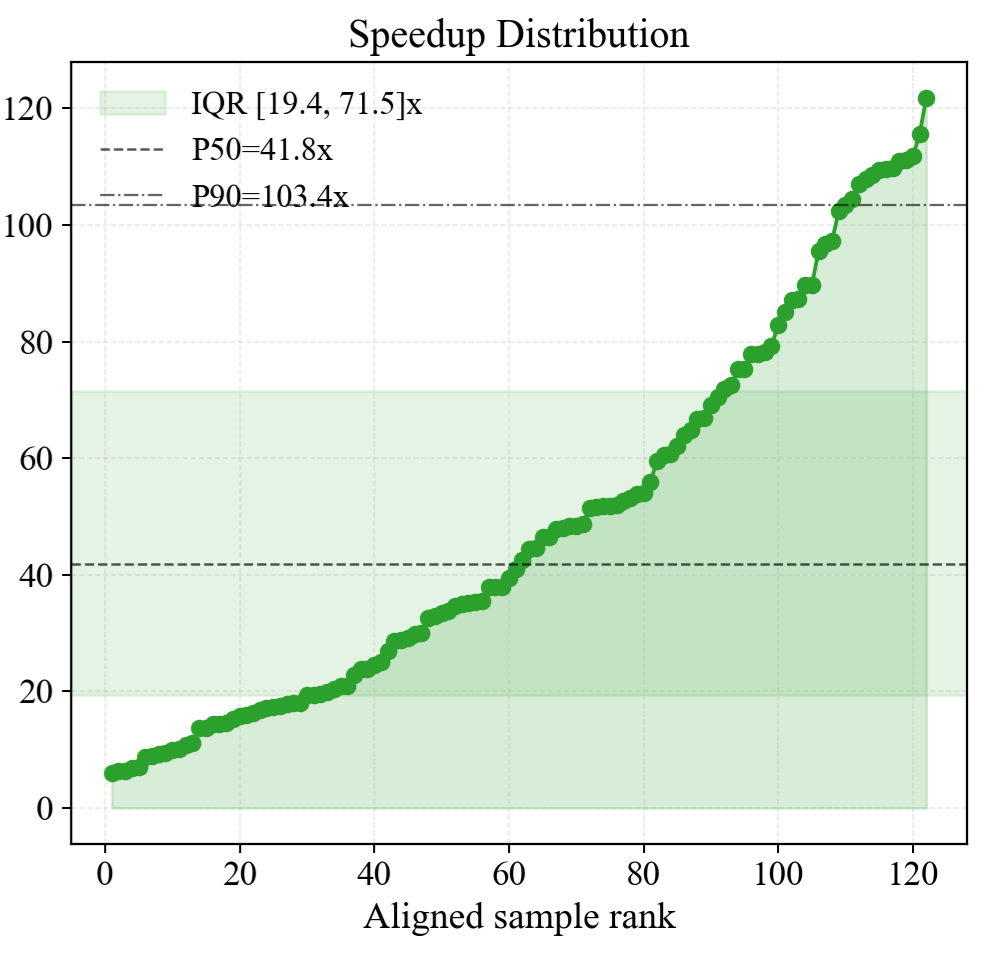}
  \\[-3pt]\scriptsize (b) Speedup
\end{minipage}
\captionsetup{font=footnotesize,skip=2pt}
\caption{Runtime comparison with different timing scopes. (a) compares DecompRL decomposition time, excluding subsequent mesh generation, with the complete measured LMFF procedure. (b) reports their runtime ratio, labeled Speedup; it is not an end-to-end meshing speedup.}
\Description{A two-part figure showing DecompRL decomposition time excluding mesh generation versus the complete measured LMFF procedure, and the corresponding runtime-ratio distribution.}
\label{fig:runtime_speedup_pair}
\end{figure}

Fig.~\ref{fig:runtime_speedup_pair} compares DecompRL decomposition time with the complete measured LMFF procedure on the same CAD models. DecompRL timing excludes subsequent hexahedral mesh generation. Figure (a) shows that decomposition times concentrate in the $10^{1}$-second range, whereas the LMFF procedure largely occupies the $10^{2}$--$10^{3}$-second range. Figure (b), labeled Speedup, reports the corresponding runtime ratio. The ranked curve remains well above 1$\times$, with a median ratio of 41.8$\times$ and a P90 ratio of 103.4$\times$; the maximum ratio reaches two orders of magnitude. These ratios compare the stated stages and are not end-to-end meshing speedups.
The LMFF comparison in this study is intended primarily as an efficiency reference. The available paired evidence does not support a dataset-wide quantitative comparison of final scaled-Jacobian distributions under identical successful-meshing conditions, so we do not infer mesh-quality superiority over LMFF from the runtime results. Quantitative SR and mesh-quality comparisons are therefore reported for the learning-based methods under the unified protocol, while LMFF is retained for runtime and representative qualitative meshing comparison.

\begin{figure}[!t]
\centering
\begin{minipage}[t]{0.41\linewidth}
  \centering
  \includegraphics[width=0.82\linewidth]{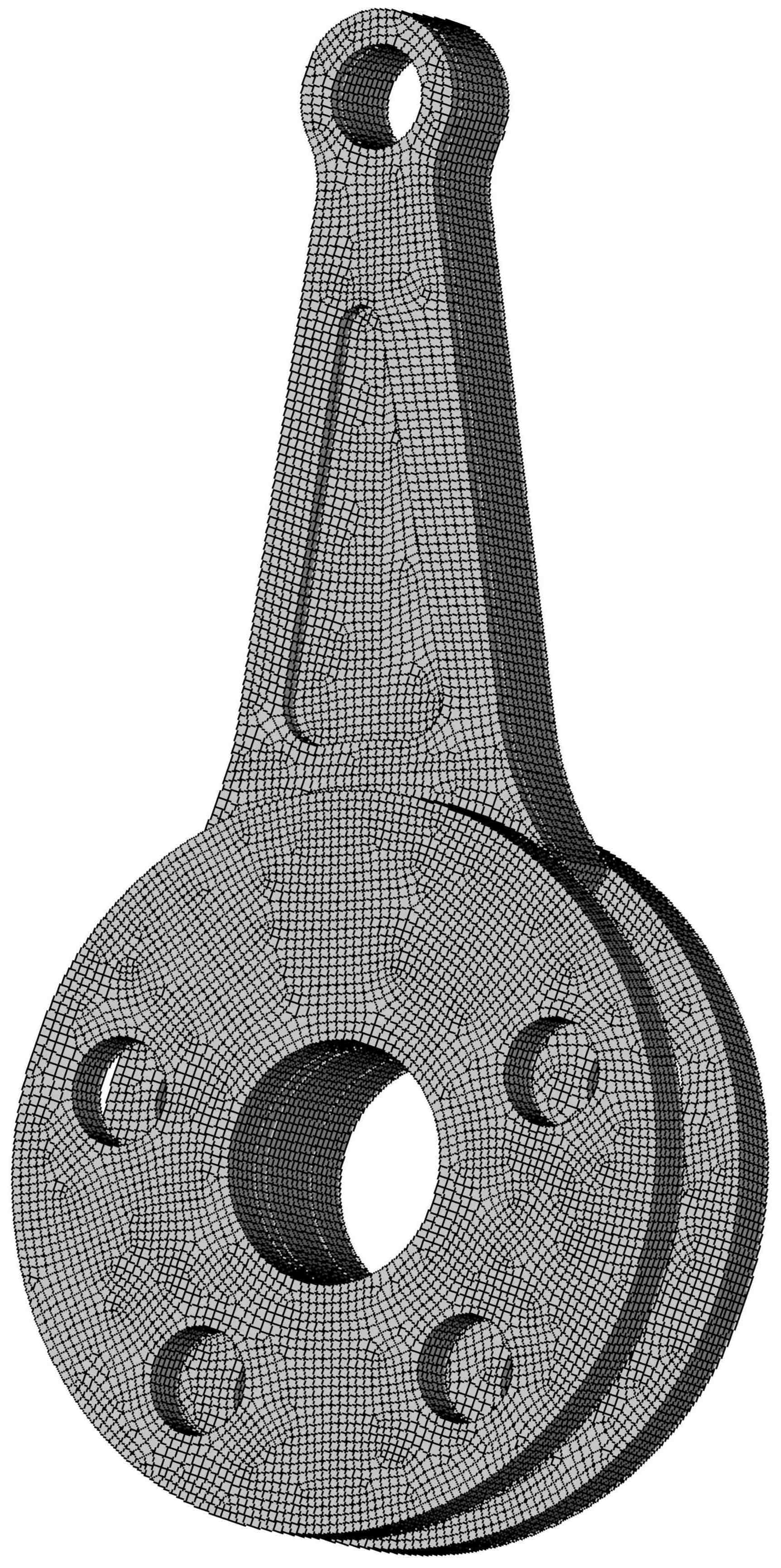}
  \\[-3pt]\scriptsize (a) Refined mesh
\end{minipage}\hspace{0.02\linewidth}
\begin{minipage}[t]{0.41\linewidth}
  \centering
  \includegraphics[width=0.82\linewidth]{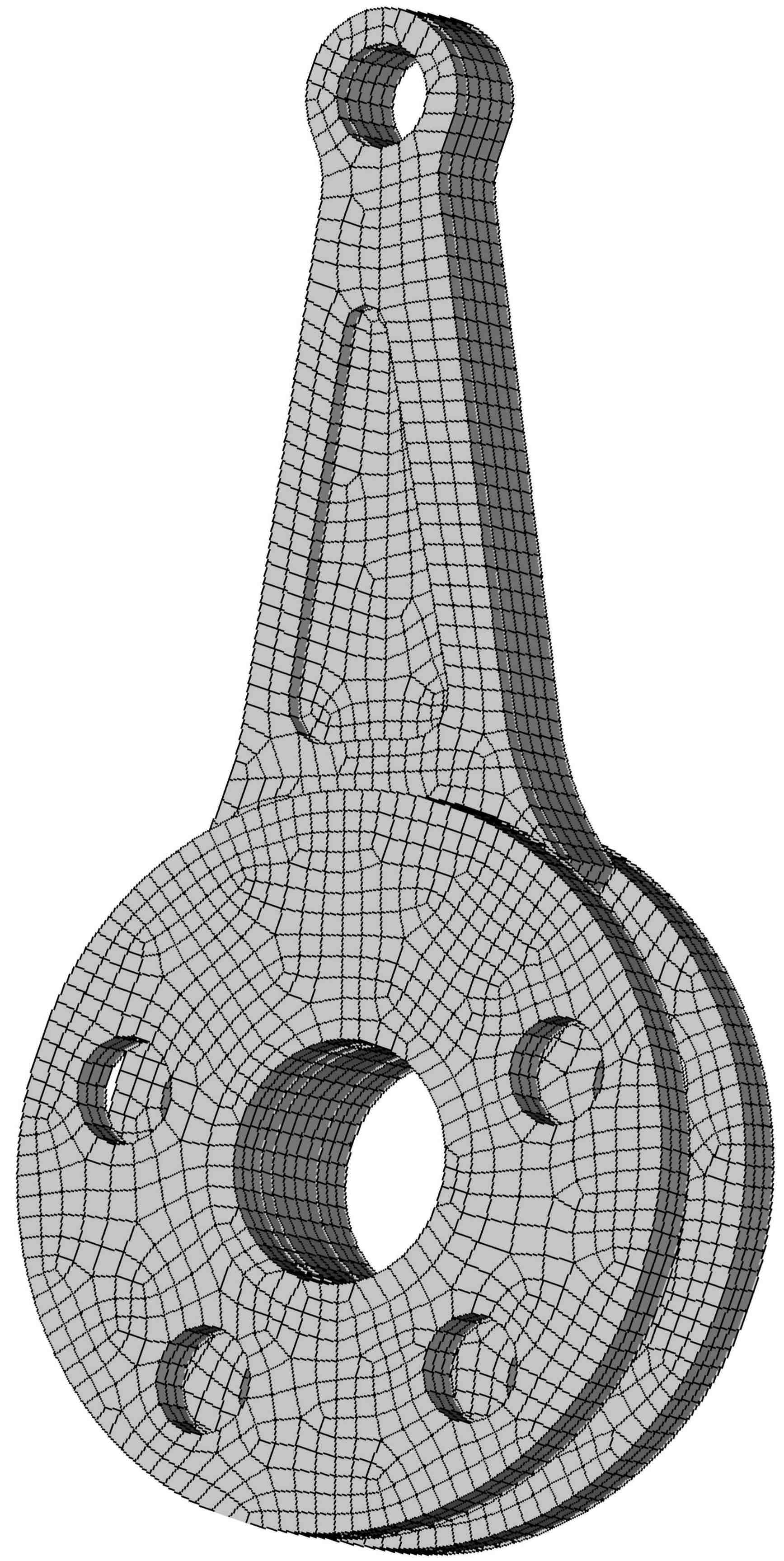}
  \\[-3pt]\scriptsize (b) Coarse mesh
\end{minipage}
\captionsetup{font=footnotesize,skip=2pt}
\caption{Refined and coarse hexahedral meshes generated from the same CAD model after decomposition.}
\Description{A two-part figure comparing a refined hexahedral mesh and a coarse hexahedral mesh on the same CAD model after decomposition.}
\label{fig:mesh_refinement_pair}
\end{figure}

Fig.~\ref{fig:mesh_refinement_pair} compares refined and coarse meshes generated from the same CAD model. Once a valid block decomposition is available, adjusting mesh resolution becomes straightforward. In contrast, frame-field-based pipelines couple mesh resolution to field construction, integer-grid-map generation, and mesh extraction, so refinement is rarely a purely local modification and often requires rerunning part of the pipeline.

Table~\ref{tab:runtime_diag} reports per-dataset decomposition diagnostics under the fixed $N=200$ setting. Here, \emph{Steps} denotes the decomposition length, \emph{Leaves} the number of exported leaf blocks, and \emph{Valid ratio} the fraction of generated candidates that pass feasibility filtering. Across ABC, Fusion360, and NIST, the policy averages roughly 5--6 steps, well below the step budget of 8, indicating that it learns to stop early rather than over-decompose. DeepCAD requires the fewest splits (1.96 on average), consistent with its simpler geometries and near-saturated SR.

\begin{table}[htbp]
\centering
\captionsetup{font=footnotesize,skip=2pt}
\caption{Per-dataset decomposition diagnostics of DecompRL under the fixed $N=200$ setting: decomposition time per sample (excluding subsequent mesh generation), decomposition length (steps), exported leaf blocks, and candidate validity ratio.}
\label{tab:runtime_diag}
\resizebox{\linewidth}{!}{%
\begin{tabular}{lcccc}
\toprule
\textbf{Dataset} & \textbf{Time/sample (s)} & \textbf{Steps} & \textbf{Leaves} & \textbf{Valid ratio} \\
\midrule
ABC       & 26.745 & 5.665 & 5.380 & 0.184 \\
Fusion360 & 24.594 & 5.770 & 5.979 & 0.190 \\
NIST      & 25.303 & 6.060 & 5.944 & 0.165 \\
DeepCAD   & 6.087  & 1.960 & 1.772 & 0.108 \\
\bottomrule
\end{tabular}
}
\end{table}

\section{Conclusion}

This paper presents DecompRL, a DRL framework for CAD model decomposition toward hexahedral meshing. Our approach combines topology-aware splitting actions with a meshability term in the reward, allowing the policy to progressively split complex CAD models into simpler sub-blocks that support hexahedral meshing. Experimental results on four public CAD benchmarks show numerically higher decomposition success rates for DecompRL on three datasets, with the largest gain on Fusion360 and more modest differences on ABC and NIST; DDPG achieves a slightly higher rate on DeepCAD. On successfully meshed outputs, DecompRL also maintains competitive hexahedral element quality.

Several limitations remain. The design of the action space is important for hexahedral decomposition, and incorporating richer action primitives may further improve the coverage of geometric configurations that can be successfully decomposed. The current action generators are primarily designed for mechanical CAD with recognizable analytic geometric structures and do not include dedicated decomposition primitives for arbitrary free-form NURBS/B-spline regions. Extending candidate generation with curvature-aware patch analysis or feature-line structures is therefore an important direction. Similarly, although the O-grid action is geometrically related to O-block layouts used in CFD, the present method does not model CFD-specific requirements such as boundary-layer growth, wall-normal alignment, or anisotropic spacing; incorporating such constraints would be required for CFD-oriented use. In addition, future work could improve training efficiency, sample efficiency, and convergence stability, especially for deeper decomposition trees and more complex CAD models.

These results should not be interpreted as a guarantee of complete, globally conforming hexahedral meshing for arbitrary CAD inputs. Local Map/Submap/Sweep compatibility does not alone establish model-level meshing success. A finite decomposition budget, early splitting decisions that restrict later feasible actions, partial geometric observations, and the limited coverage of the current action templates can all contribute to unresolved or non-meshable regions. Further work will include systematic failure analysis across geometric and topological complexity, validation of local compatibility labels against downstream meshing outcomes, and stronger resolution-aware shared-interface and interval-consistency checks.

\bibliographystyle{IEEEtran}
\bibliography{cite}

\end{document}